\documentclass[preprint,prd,aps,amssymb,nofootinbib,epsf]{revtex4}

\usepackage{bm}
\usepackage{amsmath}
\usepackage{amsfonts}
\usepackage{graphicx}
\newcommand{\be}{\begin{equation}} 
\newcommand{\ee}{\end{equation}} 
\newcommand{\bea}{\begin{eqnarray}} 
\newcommand{\eea}{\end{eqnarray}}

\newcommand{\dis}{$\displaystyle} 
\newcommand{\Lie}{\mbox{\pounds}}
\newcommand{\bsube}{\begin{subequations}}
\newcommand{\esube}{\end{subequations}}
\newcommand{\mfr}{\mathfrak r}
\newcommand{\mfro}{\mathfrak r_0}

\begin{document}  

\title{On finding fields and self-force in a gauge appropriate to separable wave equations }
\author{Tobias S. Keidl}
\affiliation{
Department of Physics, University of Wisconsin-Milwaukee, P.O. Box 413,  
Milwaukee, WI 53201}
\author{John L. Friedman} 
\affiliation{
Department of Physics, University of Wisconsin-Milwaukee, P.O. Box 413,  
Milwaukee, WI 53201}
\author{Alan G. Wiseman}
\affiliation{
Department of Physics, University of Wisconsin-Milwaukee, P.O. Box 413,  
Milwaukee, WI 53201}
%
%
%
%

\begin{abstract}
 
Gravitational waves from the inspiral of a stellar-size black hole to a
supermassive black hole can be accurately approximated by a point
particle moving in a Kerr background.  This paper presents progress on
finding the electromagnetic and gravitational field of a point particle 
in a black-hole spacetime and on computing the self-force in a ``radiation 
gauge.'' The gauge is chosen to allow one to compute the perturbed metric 
from a gauge-invariant component $\psi_0$ (or $\psi_4$) of the Weyl tensor and 
follows earlier work by Chrzanowski and Cohen and Kegeles
(we correct a minor, but propagating, error in the Cohen-Kegeles formalism).  
The electromagnetic field tensor and vector potential of a static point charge 
and the perturbed gravitational field of a static point mass in a Schwarzschild geometry are found, surprisingly, to have closed-form expressions.  The 
gravitational field of a static point charge in the Schwarzschild background
must have a strut, but $\psi_0$ and $\psi_4$ are smooth except at the particle, 
and one can find local radiation gauges for which the corresponding spin $\pm 2$ parts 
of the perturbed metric are smooth.  
Finally a method for finding the renormalized self-force from the Teukolsky equation is presented.  The method is related to the Mino, Sasaki, Tanaka and
Quinn and Wald (MiSaTaQuWa) renormalization and to 
the Detweiler-Whiting construction of the singular field. It relies on the fact 
that the renormalized $\psi_0$ (or $\psi_4$) is a {\em sourcefree} solution to the 
Teukolsky equation; and one can therefore reconstruct a nonsingular renormalized 
metric in a radiation gauge.  
\end{abstract}


\maketitle
 
\section{Introduction}
\label{intro}

Among the primary targets of the proposed space-based gravitational wave observatory, LISA, 
are waves from stellar black holes spiraling in to supermassive black holes in galactic 
centers.  Because the ratio $m/M$ is small, the orbits and gravitational waves from these binary systems can be described to high accuracy by a perturbative expansion  
with $m/M$ as the small parameter.  In addition, because tidal forces are small, 
one can  model the system as a point mass orbiting a black hole.  To zeroth orbit in $m/M$, 
the orbit is simply a geodesic of the black-hole spacetime.   To first order, the particle 
feels a self-force, whose dissipative part is the radiation-reaction force 
that drives the inspiral.  The self-force also has a conservative part that 
alters the phase of the orbit and of the emitted radiation \cite{Poisson,Pound}.  

Because the dissipative part of the self-force is antisymmetric 
under the change from ingoing to outgoing radiation -- from advanced to 
retarded fields, it can be computed from the half-retarded $-$ half-advanced 
Green's function. 
Because this Green's function is sourcefree, it is regular at the particle.  
Approximating the self-force by its dissipative 
part is an adiabatic approximation \cite{Mino1,Mino2,Mino3}, and several computations of 
orbits and waveforms have recently been carried out \cite{Drasco, DrascoHughes,
Hughes, Sago}. 

Including the conservative part of the self force is a more difficult problem, because 
it arises from a field (the half-retarded + half-advanced part of the field) that is 
singular at the particle.  One must renormalize the perturbed metric near the particle 
by subtracting off its singular part, a field singular at the position of the 
particle that does not itself contribute to the self force.  The MiSaTaQuWa prescription for 
this subtraction, given by Mino, Sasaki and Tanaka\cite{MinoSasaki} and subsequently in a 
particularly clear form by Quinn and Wald \cite{QuinnWald}, is well understood in a Lorenz 
(Hilbert, deDonder, Lorentz, harmonic) gauge. A Lorenz gauge, however, is not well-adapted 
to the Kerr geometry: Instead of the decoupled, separable Teukolsky equation that simplifies 
black-hole perturbation theory, one must solve a system of ten coupled partial differential
equations.  

We report here the beginning of a program to compute the self-force in a gauge appropriate to the 
separable wave equation, using a formalism due to Chrzanowski \cite{chrz} and to Cohen and 
Kegeles \cite{CohenKegeles,KegelesCohen} (henceforth CCK) to reconstruct the metric, in what is termed a {\em 
radiation gauge}, from either of the gauge-invariant 
components $\psi_0$ or $\psi_4$ that satisfy the Teukolsky equation\cite{TeukolskyLetter,Teukolsky}.
Cohen and Kegeles clarified and made minor corrections to Chrzanowski's work,\footnote{
 Eq. (1.7) in Ref.~\cite{chrz} appears incorrectly to identify the Hertz potential 
with a component of the perturbed Weyl tensor, but the identification is 
made correctly later in the paper, apart from a missing complex conjugation; 
Eqs. (1.3)-(1.5) seem incorrectly to imply that radiation gauges can be 
used when matter is present.  The equations are correctly given by Cohen 
and Kegeles, apart from the factor-of-two correction we make here.} 
and Wald gave a more concise derivation \cite{Wald78}. Subsequent 
work on inverting the differential equations that give the Hertz potential 
is reported in \cite{LoustoWhiting} and \cite{Ori}, and a first explicit
vacuum reconstruction of the metric in a radiation gauge is given by Yunes and 
Gonz\'alez \cite{YunesGonzalez}.  In the present paper, we give a first explicit 
reconstruction of the metric of a point particle in a radiation gauge.  
And we outline how one can use a version of the MiSaTaQuWa renormalization due to 
Detweiler and Whiting \cite{DetWhiting} to renormalize the gauge invariant component $\psi_0$ 
(or $\psi_4$) of the perturbed Weyl tensor and then to reconstruct the renormalized 
metric in a radiation gauge from the renormalized Weyl tensor. 
(A different renormalization procedure, also based on a radiation gauge, 
is given by Barack and Ori \cite{BarackOri03}.) 

Cohen and Kegeles show that the reconstruction of the vector potential from the 
spin-weight $\pm 1$ component $\phi_0$ or $\phi_2$ of the electromagnetic field tensor 
$F^{\alpha\beta}$ is closely analogous to the reconstruction of the perturbed metric 
from $\psi_0$ or $\psi_4$, and we begin with this simpler example.  
After a mathematical introduction, we obtain in Sect. \ref{charge} closed form 
expressions for all components ($\phi_0,\phi_1$ and $\phi_2$) of the electromagnetic 
field tensor of a static charge in a Schwarzschild background, a problem initially 
solved by Copson\cite{Copson}.  From $\phi_2$, we obtain the vector potential in a radiation 
gauge.  A radiation gauge exists only where there are no sources.  We observe 
that,  in a region $R$, in order to obtain from a Hertz potential a smooth vector potential in a radiation gauge, 
spheres in $R$ enclose no charge; we conjecture that this is generally true.  One can obtain a smooth vector potential in a radiation gauge by adding an 
$l=0$ part of the field in another gauge. 
(And, at least in a Schwarzschild background, 
one can alternatively add an $l=0$ field in a radiation gauge that does not arise from a Hertz potential.)   
 
We next (in Sect. \ref{mass1}) obtained a closed-form expression for the components 
$\psi_0$ and $\psi_4$ of the perturbed Weyl tensor of a static point mass 
in a Schwarzschild background.  
There is no consistent solution to the perturbed Einstein equation whose source is 
a static point mass: The mass must be supported.  The support, however, can be a 
strut that does not contribute to the spin-weight $\pm 2$ components of the Weyl 
tensor.  In reconstructing the metric, we again find that the perturbed metric 
in a radiation gauge that arises from a Hertz potential is smooth only if spheres enclose no perturbed mass.  
Again one can obtain a smooth perturbed metric in a radiation gauge by adding an 
$l=0$ part of the field in another gauge.  This is part of the freedom one has 
to add an algebraically special perturbation without changing the spin-weight 
$\pm 2$ part of the field on which the radiation gauge is based. Regardless of 
the choice of gauge, one cannot 
simultaneously make the metric smooth everywhere inside and outside the radius 
of the point mass. To obtain a smooth metric in a radiation gauge for the 
spin $\pm 2$ part of the metric, one can 
add a strut by adding an algebraically special perturbation -- 
in this case a perturbed C metric having a nonzero deficit angle along an axis 
through the particle.   

The strut is a feature of our static example that does not appear in an inspiral 
problem, and we conclude our sample reconstructions by considering a point mass 
in flat space, in which we choose a null tetrad that is again not centered at 
the perturbing mass.  The formalism is now very close to that of the point 
charge in a Schwarzschild background, in which the only singularity in the 
perturbed metric outside the particle is pure gauge and can be removed by adding an $l=0$ perturbation in a different gauge.  

Finally, in Sect.~\ref{renormalization}, we note that the form given by Detweiler 
and Whiting for the singular part of the perturbed metric can be used to find 
the singular part of $\psi_0$ (or $\psi_4$).  The renormalized Weyl tensor component, 
$\psi^{\rm ren} = \psi^{\rm ret}_0 -  \psi^{\rm sing}_0$, satisfies the 
{\em sourcefree} Teukolsky equation.  As a result, one can find a nonsingular 
renormalized metric perturbation in a radiation gauge.  The motion of a particle 
is then given to first order in $m/M$ by the requirement that 
it move on a geodesic of the renormalized metric.

\section{Mathematical Preliminaries}
\label{preliminaries}
Greek letters early in the alphabet $\alpha, \beta, \ldots$ will be abstract spacetime 
indices; letters $\mu, \nu, \ldots$ will be concrete indices, labeling components 
along the tetrad defined in Eq.~(\ref{tetrad}) below. 
We adopt the $+---$ signature of Newman and Penrose (NP)\cite{NewmanPenrose1,NewmanPenrose2}, writing the Schwarzschild metric in the form
\begin{eqnarray}
ds^2= \frac{\Delta}{r^2}dt^2-\frac{r^2}{\Delta}dr^2
     -r^2d\theta^2-r^2\sin^2\theta d\phi^2,\qquad \qquad 
     \mbox{ where } \Delta=r^2-2Mr. 
\end{eqnarray}
We primarily use the Kinnersley tetrad\cite{Kinnersley}, $\{\bf e_\mu\}$ (numbered following NP notation),
\begin{eqnarray}
e_1^\alpha\equiv l^\alpha &=& \frac{r^2}{\Delta}t^\alpha+r^\alpha 
\qquad\qquad
e_2^\alpha\equiv n^\alpha =\frac{1}{2}t^\alpha-\frac{1}{2}\frac{\Delta}{r^2}r^\alpha
\nonumber\\ 
e_3^\alpha\equiv  m^\alpha&=& \frac{1}{\sqrt{2}}(\hat{\theta}^\alpha+i\hat{\phi}^\alpha)
\qquad 
e_4^\alpha\equiv {\overline m}^\alpha,
\label{tetrad}
\end{eqnarray}
where we denote by $t^\alpha$ and $r^\alpha$ the vectors ${\bm\partial}_t$ and
${\bm\partial}_r$ and by $\hat\theta^\alpha$ and $\hat\phi^\alpha$ the unit 
vectors \dis\frac1r{\bm\partial}_\theta$ and \dis\frac1{r\sin\theta}{\bm\partial}_\phi$. 
The derivative operators associated with $l^\alpha, n^\alpha, m^\alpha$ and $\overline m^\alpha$
are, as usual denoted by $D, {\mathbf\Delta}, \delta$ and $\bar \delta$, respectively, but 
with a boldface $\bm \Delta$ to distinguish this symbol from $\Delta=r^2-2Mr$. 
 
In terms of the nonzero spin coefficients of the Kinnersley tetrad, 
\begin{eqnarray}
\rho=-\frac{1}{r} \qquad \qquad \beta=-\alpha=\frac{\cot\theta}{2\sqrt{2}r} \qquad \qquad 
\gamma=\frac{M}{2r^2} \qquad \qquad \mu=-\frac{1}{2r} \frac{\Delta}{r^2},
\label{spincoefficients}\end{eqnarray} 
the corresponding nonzero Christoffel symbols have the form 
\begin{displaymath}
\begin{array}{l l l l} \Gamma^1_{\ 12}=2\gamma\quad\qquad 
& \Gamma^2_{\ 22}=-2\gamma\quad\qquad & \Gamma^3_{\ 33}=2\beta\quad\qquad 
& \Gamma^1_{\ 43}=\mu  \\ \\  
\Gamma^3_{\ 13}=-\rho & \Gamma^3_{\ 23}=\mu 
& \Gamma^1_{\ 34}=\mu & \Gamma^2_{\ 43}=-\rho  \\ \\   
\Gamma^4_{\ 14}=-\rho & \Gamma^4_{\ 24}=\mu & \Gamma^2_{\ 34}=-\rho 
& \Gamma^4_{\ 43}=-2\beta  \\ \\  
& & \Gamma^3_{\ 34}=-2\beta & \Gamma^4_{\ 44}=2\beta. 
\end{array}
\end{displaymath} 
Here, for example, 
\dis\Gamma^\mu_{\ 12}e_\mu{}^\alpha\equiv e_2{}^\beta \nabla_\beta\, e_1{}^\alpha$.

The electromagnetic field $F_{\alpha\beta}$ has independent complex components, 
\bea
\phi_0&=&F_{\alpha\beta}l^\alpha m^\beta \label{em_phi0_def}\\
\phi_1&=&\frac{1}{2}F_{\alpha\beta}(l^\alpha n^\beta-m^\alpha{\overline m}^\beta) \label{em_phi1_def}\\
\phi_2&=&F_{\alpha\beta}{\overline m}^\alpha n^\beta,  \label{em_phi2_def}
\eea
with spin-weights 1, 0, and -1, respectively, where each occurrence of 
$m^\alpha$ ($\overline m^\alpha$) contributes 1 (-1) to the spin-weight. Similarly, the 
Weyl tensor $C_{\alpha\beta\gamma\delta}$ has independent components   
\begin{eqnarray}\label{weyl}
\Psi_0&=&-C_{\alpha \beta \gamma \delta} l^\alpha m^\beta l^\gamma m^\delta \nonumber \\
\Psi_1&=&-C_{\alpha \beta \gamma \delta} l^\alpha n^\beta l^\gamma m^\delta \nonumber \\
\Psi_2&=&-C_{\alpha \beta \gamma \delta} l^\alpha m^\beta {\overline m}^\gamma n^\delta \nonumber \\
\Psi_3&=&-C_{\alpha \beta \gamma \delta} l^\alpha n^\beta {\overline m}^\gamma n^\delta \nonumber \\
\Psi_4&=&-C_{\alpha \beta \gamma \delta} n^\alpha {\overline m}^\beta n^\gamma {\overline m}^\delta,
\label{Psi_i}\end{eqnarray}
with spin-weights $2,1,0,-1$ and $-2$, respectively. 

In the Schwarzschild geometry, only $\Psi_2$ is nonzero, and it has value 
\be
\Psi_2 = -\frac M{r^3}.
\ee 

We define the perturbation $h_{\alpha\beta}$ of a background metric $g_{\alpha\beta}$ by considering a family of metrics
$g_{\alpha\beta}(\zeta)$, and writing 
\be
h_{\alpha\beta} =\left. \frac d{d\zeta} g_{\alpha\beta}\right|_{\zeta=0}.
\ee
The corresponding components of the perturbed Weyl tensor along the unperturbed  
tetrad will be denoted by the lower-case symbols, $\psi_0, \ldots, \psi_4$: For example,
\be
\psi_3 = -\left.\frac d{d\zeta} C_{\alpha\beta\gamma\delta}\right|_{\zeta=0}
	l^\alpha n^\beta {\overline m}^\gamma n^\delta.
\ee  

  Tensor components of the Kinnersley tetrad with spin-weight $s$ have angular behavior given by spin-weighted spherical harmonics, 
${}_sY_{lm}(\theta,\phi)$, when the tensor belongs to an $l,m$ representation of 
the rotation group \cite{GoldbergEtal}.  To define ${}_sY_{lm}$, one first introduces operators $\eth$ (edth) and $\bar\eth$ that respectively raise and lower by 1 the spin-weight of a quantity $\eta$ of spin-weight $s$: 
\begin{eqnarray}
\eth\eta&=&-(\sin\theta)^s\left(\partial_\theta+i \csc\theta \partial_\phi\right)(\sin\theta)^{-s}\eta \label{green_eth_a}\\
&=&-\left(\partial_\theta+i\csc\theta\partial_\phi-s\cot\theta\right)\eta, \label{green_eth_b}
\end{eqnarray}
\begin{eqnarray}
\bar{\eth}\eta&=&-(\sin\theta)^{-s}\left(\partial_\theta-i\csc\theta\partial_\phi\right)(\sin\theta)^s\eta \label{green_eth_bar_a} \\
&=&-\left(\partial_\theta-i\csc\theta\partial_\phi+s\cot\theta\right)\eta. \label{green_eth_bar_b}
\end{eqnarray}
Then, for each value of $s$, the spin-weighted spherical harmonics are a complete set of 
orthonormal functions on the two-sphere, given by
\be
{}_sY_{l m}= \left\{\begin{array}{ll}
              \left[ (l-s)!/(l+s)! \right]^{1/2}\eth^s Y_{l m},& 0\le s\le l,
              \\
              (-1)^s\left[(l+s)!/(l-s)!\right]^{1/2}\bar{\eth}^{-s} Y_{l m},&
	 -l\le s\le 0.
	\end{array}\right.
 \label{green_sylm_a}\ee
They satisfy the identities, 
\begin{eqnarray}
{}_s\!Y_{l m}^*&=& (-1)^{m+s} \,_{-s}Y_{l m}, 
\label{green_sylm_prop_1} \\
\eth\ _s\!Y_{l m}&=& \left[(l-s)(l+s+1)\right]^{1/2}{}_{s+1}Y_{l m},
\label{green_sylm_prop_2} \\
\bar{\eth}\ _s\!Y_{l m}&=& -\left[(l+s)(l-s+1)\right]^{1/2}{}_{s-1}Y_{l m}, 
\label{green_sylm_prop_3} \\
\bar{\eth}\eth\ _s\!Y_{l m}&=&-(l-s)(l+s+1)\ _sY_{l m}, 
\label{green_sylm_prop_4}\\
\sum_{l=\left|s\right|}^\infty \sum_{m=-1}^l{}_sY_{l m}^*(\theta',\phi')\ _sY_{lm}(\theta,\phi)
&=&\delta(\cos\theta'-\cos\theta)\delta(\phi'-\phi), 
\label{green_sylm_prop_5}\\
\int d\Omega \ _s\!Y^*_{l m}\ _s\!Y_{l'm'}&=&\delta_{ll'}\delta_{mm'}.
\label{green_sylm_prop_6}
\end{eqnarray}

The spin $\pm1$ components $\phi_0$ and $\phi_2$ of the electromagnetic field and the 
spin $\pm2$ components $\psi_0$ and $\psi_4$ of the perturbed Weyl tensor satisfy decoupled wave equations, 
namely the Bardeen-Press equation \cite{BardeenPress} and its electromagnetic analog. 
These are the a=0, spin $\pm1$ and spin $\pm2$ cases of the Teukolsky equation\cite{TeukolskyLetter, Teukolsky},\footnote{Teukolsky's expressions for the $s=\pm2$ Schwarzschild source functions in Ref.~\cite{TeukolskyLetter} differ from those in Ref.~\cite{Teukolsky} by an overall sign.  With our conventions, the signs agree with Ref.~\cite{Teukolsky}.} and 
they have the form
\begin{eqnarray}
\left[\frac{r^2}\Delta\partial_t^2 -2s\left(\frac M\Delta - \frac1r\right)\partial_t 
+\mathbb{L}\right]\psi = 4\pi T, 
\label{bp}
\end{eqnarray}
with 
\be
\mathbb{L} \equiv -\frac{\Delta^{-s}}{r^2} \frac{\partial}{\partial r} 
\left( \Delta^{s+1} \frac{\partial} {\partial r} \right) 
-\frac{1}{r^2} \bar{\eth}\eth,
\label{L}
\ee
where $\psi$ is any of the functions listed in the first column of Table \ref{sourcetable}
and $T$ is the corresponding source listed in the third column.  The source involves 
components of the current four-vector $J^\alpha$ for electromagnetism and of the 
stress-energy tensor $T^{\alpha\beta}$ for gravity.

\begin{table}[h]
\caption{Source function}
\label{sourcetable}
\begin{center}
\begin{tabular}{|c|c|c|}
\hline
$\psi$ & $\ \ s$ & $T$\\
\hline
$\phi_0$ & $\ \ 1$ & $\delta J_1-(D-3\rho)J_3$\\
\hline
$\rho^{-2}\phi_2$ & $-1$ & $\rho^{-2}\left\{(\Delta +3\mu)J_4
	-\bar{\delta}J_2\right\}$\\
\hline
$\psi_0$ & $\ \ 2$ & $2 (\delta -2\beta)[(D-2\rho)T_{13}-\delta T_{11}]+$\\
& & $2(D-5\rho)[(\delta-2\beta)T_{13}-(D-\rho)T_{33}]$\\
\hline
$\rho^{-4}\psi_4$ & $-2$ & $2 \rho^{-4}(\Delta +2\gamma+5\mu)
  [(\bar{\delta}-2\beta)T_{2 4}-(\Delta+\mu)T_{44}]+$ \\
& & $2 \rho^{-4}(\bar{\delta}-2\beta)[\Delta+2\gamma+2\mu)T_{24}-\bar{\delta}T_{22}]$\\
\hline
\end{tabular}
\end{center}
\end{table}
\newpage

\noindent
{\em Static Green's function}

To compute the fields of static sources, we use the time-independent 
forms of these equations, 
\be 
\mathbb{L}\psi = 4\pi T.
\label{static-bp}\ee

Let $x$ and $x'$ denote points of a $t=$const hypersurface of the Schwarzschild 
geometry.  
We find below that the source terms for a static particle are related by the 
operator $\eth$ to a point source of the form $\delta^3(x,x')$, normalized by \\
\be
 1=\int dV'\, \delta^3(x,x') = \int d^3x' \sqrt{{}^3g(x')}\ \delta^3(x,x').
\label{deltafn}
\ee
In Schwarzschild coordinates, with $\sqrt{^3g} = r^3\Delta^{-1/2} \sin\theta$, we have
\be
 \delta^3(x,x')=r^{-3}\Delta^{1/2}\delta(r-r')\delta(\cos\theta-\cos\theta')\delta(\phi-\phi').
\ee
Denote by $G(x,x')$ the Green's function satisfying Eq.~(\ref{static-bp}) with this source,  
\be
\mathbb{L}G(x,x')=-4\pi\delta^3(x,x')\label{green_definitions_a}.
\ee
A solution $\psi(x)$ to Eq.~(\ref{static-bp}) then has the form
\be
\psi(x)=\int G(x,x')T(x')dV'\label{green_definitions_b}.
\ee

We construct $G(x,x')$ from solutions to the homogenous equation $\mathbb L\psi=0$.
From Eq.~(\ref{L}), these have the form  $R(r)S(\theta,\phi)$, with  
\be
    \bar\eth\eth S = \lambda S,\qquad
     \frac{\Delta^{-s}}{r^2}\frac{\partial}{\partial r} 
\left(\Delta^{s+1}\frac{\partial R} {\partial r} \right)+\frac\lambda{r^2} R=0.
\ee
From Eq.~(\ref{green_sylm_prop_4}), the solutions to the angular equation 
are the spin-weighted spherical harmonics with eigenvalues 
$\lambda=-(l-s)(l+s+1)$.  The radial equation then takes the form
\begin{eqnarray}\label{green_radial}
\frac{\Delta^{-s}}{r^2}\frac{\partial}{\partial r} 
\left(\Delta^{s+1}\frac{\partial R} {\partial r} \right)
-\frac{1}{r^2}(l-s)(l+s+1)R = 0.  
\end{eqnarray}
Its solutions are given in terms of the associated Legendre polynomials by 
\begin{equation}\label{radial_solution}
R(r)= \frac{a}{\Delta^{s/2}}P_l^s\left(\frac{{\mathfrak r}}{M}\right)
     +\frac{b}{\Delta^{s/2}}Q_l^s\left(\frac{{\mathfrak r}}{M}\right),
\end{equation}
where $a$ and $b$ are arbitrary constants and ${\mathfrak r} \equiv r-M$.

Using the Wronskian
\begin{equation}\label{wronskianpq}
W\left[P^s_l(z), Q^s_l(z)\right] = \frac{(-1)^s(l+s)!}{(1-z^2)(l-s)!},
\end{equation}
for integral values of $l$ and $s$, we obtain the Wronskian of the two radial solutions,
\begin{equation}\label{wronskian}
W\left[\frac{1}{\Delta^{s/2}}P_l^s\left(\frac{{\mathfrak r}}{M}\right), 
\frac{1}{\Delta^{s/2}}Q_l^s\left(\frac{{\mathfrak r}}{M}\right)\right] = (-1)^{s+1}\frac {M^2}{\Delta^{s+1}}\frac{(l+s)!}{(l-s)!}.
\end{equation}
Using this relation and the completeness (\ref{green_sylm_prop_5}) and orthogonality (\ref{green_sylm_prop_6}) relations for the ${}_s\!Y_{lm}$ to compute 
for each $l,m$ the discontinuity in Eq.~(\ref{green_definitions_a}) across $r=r'$, we find for the Green's function the form

\begin{equation}\label{green_greens_function}
G( x, x') = (-1)^s \frac{ 4 \pi}{M}\left(\frac{\Delta'}\Delta \right)^{s/2}\sum_{l=\left|s\right|} ^{\infty}\sum_{m=-l} ^l \frac{(l-\left|s\right|)!} {(l+\left|s\right|)!} P_l ^{\left|s\right|} \left( \frac{{\mathfrak r}_<}{M} \right) Q_l ^{\left|s\right|} \left( \frac{{\mathfrak r}_> }{M}\right)\ _sY _{l m}(\theta, \phi)\ _sY^* _{l m} (\theta', \phi'),
\end{equation}
where ${\mathfrak r}_<\equiv\min(r,r')-M$ and ${\mathfrak r}_>\equiv\max(r,r')-M$.

\subsection{Tetrads smooth at $\theta=0$ and $\theta=\pi$ and a criterion for smoothness 
of a tensor field.}

The Kinnersley tetrad is singular when $\theta=0$ and $\theta=\pi$.  
In order to disentangle (1) the singularity arising from the choice of tetrad from 
(2) a singularity arising from the use of the radiation gauge and 
(3) a physical singularity associated with a static particle on a Schwarzschild background, 
we will use, in addition to the Kinnersley tetrad, a closely related tetrad, 
$\{{\bf e}^+_{\mu}\}=\{l^\alpha, n^\alpha, m_+^\alpha=e^{i\phi}m^\alpha , \overline m_+^\alpha=e^{- i\phi}\overline m^\alpha\}$, 
that is smooth everywhere except $\theta =\pi$. (By {\em smooth} we 
always mean $C^\infty$).
One can similarly replace $m^\alpha$ by $e^{-i\phi}m^\alpha $ to obtain a null tetrad $\{{\bf e}^-_{\mu}\}$, smooth 
everywhere except $\theta = 0$.  Because there is no continuous vector field nonzero everywhere on a two-sphere, and because $m^\alpha$ must be tangent to the symmetry spheres if it is orthogonal to the 
principal null directions, no null tetrad based on the principal null directions is smooth everywhere. 

After showing that $e^{\pm i\phi}m^\alpha $ is smooth near $\theta=0$ ($\theta = \pi$), we will find a 
simple criterion for smoothness of a tensor field near $\theta = 0$ and $\theta=\pi$ that 
involves only its components in the Kinnersley tetrad.   

To see that $e^{i\phi}m^\alpha $ is smooth near $\theta=0$, we introduce a chart $\{t,X,Y,Z\}$ that is smooth 
on the spacetime exterior to the horizon: With
\be 
 X= r\sin\theta\cos\phi, \quad Y= r\sin\theta\sin\phi, \quad Z= r\cos\theta,
\label{xyz}
\ee
the Schwarzschild metric has the form 
\[
 ds^2 = \left(1-\frac{2M}r\right)dt^2 - dX^2-dY^2-dZ^2 -\frac{2M}{r^2(r-2M)}(XdX+YdY+ZdZ)^2,
\]
smooth for $r>2M$.  Then $e^{i\phi}m^\alpha $ is smooth, because its components in this Cartesian chart are 
smooth.  We have 
\bsube
\bea
     m'^Z&=&e^{i\phi}m^Z = e^{i\phi} m^\theta\partial_\theta Z= -\frac1{\sqrt2}\frac{X+iY}r,\\
     m'^X &=& e^{i\phi}m^X 
	= e^{i\phi}\frac1{\sqrt2\,r} \left(\partial_\theta +\frac i{\sin\theta} \partial_\phi\right) X
	= \frac1{\sqrt2}\left[1-\frac Xr(X+iY)\frac{r-Z}{X^2+Y^2}\right],\\
 m'^Y &=& e^{i\phi}m^X 
	= e^{i\phi} \frac1{\sqrt2\,r}\left(\partial_\theta +\frac i{\sin\theta}\partial_\phi\right) Y 
	= \frac1{\sqrt2}\left[i-\frac Yr(X+iY)\frac{r-Z}{X^2+Y^2}\right].
\eea \esube
Finally, \dis \frac {r-Z}{X^2+Y^2}$ is smooth everywhere except the negative Z-axis, 
because it is analytic in $\{X,Y,Z\}$: This is obvious for $(X,Y)\neq 0$. To show that 
it is true for $(X,Y)=0,\quad Z>0$, write $w=(X/r)^2+(Y/r)^2$. Then, for $Z> 0$,  
\be
 F(w)\equiv r\frac{r-Z}{X^2+Y^2} = \frac{1-\sqrt{1-w}}w = \sum a_n w^n, \qquad a_n 
	= \frac12\frac{(1-\frac12)}2\frac{(2-\frac12)}3\cdots \frac{(n-\frac12)}{n+1} < 1,
\ee
implying $F(w)$ analytic for $|w|<1$. Thus 
$\{l^\alpha,n^\alpha,  m_+^\alpha,\bar m_+^\alpha\}$ is a smooth (and analytic) basis 
for $r>2M$, except on the negative $Z$-axis.  Similarly, 
$\{l^\alpha,n^\alpha,  m_-^\alpha, \bar m_-^\alpha\}$ is a smooth (and analytic) basis 
for $r>2M$, except on the positive $Z$-axis.

A tensor field is smooth if and only if its components in a smooth basis 
are smooth. Thus a tensor field is smooth on the positive  (negative)
Z-axis if and only if its components along the basis $\{{\bf e}^\pm_\mu\}$
are smooth. Now a function $f e^{im\phi}$ (with $f$ independent of $\phi$) 
is smooth at $\theta=0$ (and 
$f e^{-im\phi}$ is smooth at $\theta=\pi$) if and only if 
$f = g \sin^m\theta$, with $g$ smooth.  If $f$ is a spin-weight $m$ 
component of a tensor in the Kinnersley basis, then $f e^{\pm im\phi}$ is 
the corresponding component in the basis $\{\bf e^\pm_\mu\}$.  
Replacing $\sin^m\theta$ by $\theta^m$ for $\theta$ near $0$ and $(\pi-\theta)^m$ for $\theta$ near $\pi$, we obtain the following criterion for the smoothness of a tensor:\\ 
{\em Proposition}.  A tensor 
is smooth near $\theta=0$ if and only if its components with spin-weight $m$ in a Kinnersley tetrad have the form $\theta^m g$, with $g$ smooth near 
$\theta =0$; and it is smooth near $\theta=\pi$ if and only if its 
components in a Kinnersley tetrad have the form $(\pi-\theta)^m g$, with 
$g$ a function smooth near $\theta=\pi$.  \\

In the next three sections, we will use the CCK formalism to write 
the vector potential $A_\alpha$ in terms of a scalar $\Phi$ and 
to write the perturbed metric $h_{\alpha\beta}$ in terms of a 
scalar $\Psi$, a Hertz potential.  We show in Appendix \ref{appendixa} 
that the Hertz potentials $\Psi$ and $\Phi$ associated with a Kinnersley 
tetrad are related by phases, $e^{-2i\phi}$ and $e^{-i\phi}$ respectively,
to the Hertz potential associated with a smooth basis. It follows that 
$h_{\alpha\beta}$ and $A_\alpha$ are smooth if $\Psi/\sin^2\theta \ $ 
and $\Phi/\sin\theta $ are smooth. (As noted below, the converse 
is not true: A singular Hertz potential can yield a smooth $h$ or $A$.)

\section{Static Charge in a Schwarzschild Spacetime}
\label{charge}

\subsection{Outline}

We can now compute the electromagnetic field of a static point charge in a Schwarzschild background, 
finding its complex components (\ref{em_phi0_def}-\ref{em_phi2_def}) along the Kinnersley tetrad. 
We first use the decoupled equation (\ref{bp}) to find $\phi_2$, obtaining a 
closed-form expression.  Then, following Cohen and Kegeles 
\cite{CohenKegeles,KegelesCohen}, we construct a 
potential $\Phi$, in terms of which we compute a vector potential $A_\alpha$ and the remaining components 
$\phi_0$ and $\phi_1$ of the field.\footnote{
Cohen and Kegeles denote by the same symbol $\Psi$ Hertz potentials both for
the electromagnetic field and for the perturbed Weyl tensor.  
To avoid this ambiguity, we denote by $\Phi$ 
the electromagnetic Hertz potential, retaining $\Psi$ for the gravitational Hertz potential:
$\Phi = \Psi$(Cohen-Kegeles, electromagnetic), $\Psi = \Psi$(Cohen-Kegeles, gravitational).
}    
Because $\phi_2$ has spin-weight -1, it has no $l=0$ part, and the monopole part of the field -- 
a change in the charge of the black hole -- 
must be determined separately.  In this formalism,
the potential $\Phi$ is obtained from $\phi_2$ by solving a second-order differential equation, and 
freedom to add electric and magnetic charge to the black hole can be identified with one of 
the constants of integration.   
Because the gauges for $A_\alpha$ introduced by Cohen and Kegeles exist globablly only for a source-free solution, 
we find local gauges that are singular on different parts of the axis of symmetry.  
Explicit gauge transformations are found relating 
the vector potential to that found by Copson in 1928 \cite{Copson}.

\subsection{Computing $\phi_2$ from the decoupled wave equation}

Consider a static charge $e$ at a point with radial coordinate $r_0$, outside the horizon of a Schwarzschild black hole; and choose the $\theta=0$ axis to pass through the charge.  The charge 
has current 4-vector $J^\alpha=\rho_e u^\alpha$, with 4-velocity 
\dis u^\alpha = \frac r{\Delta^{1/2}} t^\alpha$ and with charge density given in the notation 
of Eq.~(\ref{deltafn}) by
\be 
  \rho_e(x) = e \delta^3(x,x_0) = e\frac{\Delta^{1/2}}{r^3}\delta(r-r_0)\delta(\cos\theta-1)\delta(\phi).
\ee

The component $\phi_2$ satisfies the wave equation (\ref{bp}) with spin $s=-1$.  Note first that 
in the source term corresponding to $s=-1$ in Table \ref{sourcetable} the angular component 
$J_4$ vanishes. The only contribution to $T$ is then from the component 
\dis J_2 \equiv J_\alpha n^\alpha$, 
\be T = -\frac1{\rho^2} \bar\delta J_2.
\ee
Acting on the spin-weight 0 component $J_2$, the operator $\bar\delta$ is 
\dis-\frac{1}{r\sqrt{2}}\bar{\eth}$, implying
\be\label{em_source}
T=\frac{r}{\sqrt{2}}\bar{\eth}J_2.
\ee
From Eq.~(\ref{tetrad}), we have 
\bea 
J_2 &=& \left(\rho_e \frac r{\Delta^{1/2}}t_\alpha \right)\left( \frac12 t^\alpha \right)
    = -\frac12 \frac{\Delta^{1/2}}r \ \rho_e \nonumber\\ 
  &=& -\frac12e\frac\Delta{r^4}\delta(r-r_0)\delta(\cos\theta-1)\delta(\phi).
\eea  

We next decompose the source into spin-weighted spherical harmonics, using the completeness 
relation (\ref{green_sylm_prop_5}):
\begin{eqnarray}
J_2 &=&-\frac12e\frac\Delta{r^4}\delta(r-r_0)
	\sum_{l=0}^\infty\sum_{m=-l}^l Y_{l m}(\theta,\phi)Y_{l m}^*(0,0),\\
T(x)&=& -\frac{e\Delta}{2\sqrt{2}r^3}\delta(r-r_0)
        \bar{\eth}\sum_{l=0}^\infty\sum_{m=-l}^l Y_{l m}(\theta,\phi)Y_{l m}^*(0,0)
\label{em_source_projected_2}\nonumber\\
&=&-\frac{e\Delta}{2\sqrt{2}r^3}\delta(r-r_0)
	\sum_{l=1}^\infty\sum_{m=-l}^l[l(l+1)]^{1/2}\ _{-1} Y_{l m}(\theta,\phi)Y_{l m}^*(0,0).
\label{em_source_projected_3}
\end{eqnarray}

It is now straightforward to compute $\phi_2$ from the Green's function $G(x,x')$ of 
Eq.~(\ref{green_greens_function}), using orthonormality (\ref{green_sylm_prop_6}) of the ${}_sY_{lm}$.
We have 

\begin{eqnarray}
\rho^{-2}\phi_2&=&\int G(x,x')T(x')\sqrt{-g}d^3x' \label{em_phi2_integral_a}
\nonumber\\
&=&\int\left[-\frac{4\pi}{M}\left(\frac{\Delta}{\Delta'}\right)^{1/2}\sum_{l=1}^\infty \sum_{m=-l}^l\frac{1}{l(l+1)}P_l^1\left(\frac{{\mathfrak r}_<}{M}\right)Q_l^1\left(\frac{{\mathfrak r}_>}{M}\right)\ _{-1}Y_{l m}(\theta,\phi)\ _{-1}\!Y_{l m}^*(\theta',\phi')\right] \nonumber\\
& &\times \left[-\frac{e\Delta'}{2\sqrt{2}r'^3}\delta(r'-r_0)\sum_{n=1}^\infty \sum_{p=-n}^n[n(n+1)]^{1/2}\ _{-1}\!Y_{n p}(\theta',\phi')Y_{n p}^*(0,0)\right]r'^2dr'd\Omega' \label{em_phi2_integral_b}
\nonumber\\
&=&\frac{4\pi e (\Delta \Delta_0)^{1/2}}{2\sqrt{2}Mr_0}\sum_{l=1}^\infty\sum_{m=-l}^l\frac{1}{[l(l+1)]^{1/2}}P^1_l\left(\frac{{\mathfrak r}_<}{M}\right)Q^1_l\left(\frac{{\mathfrak r}_>}{M}\right)
	{}_{-1}\!Y_{l m}(\theta,\phi)Y^*_{l m}(0,0), 
\label{em_phi2_mode}
\end{eqnarray}
where ${\mathfrak r}_<\equiv \min(r,r_0)-M$, ${\mathfrak r}_>\equiv \max(r,r_0)-M$ and $\Delta_0=r_0^2-2M r_0$.  Summing this series (details in the Appendix \ref{appendixb}) yields the simple closed-form expression,
\begin{equation}\label{em_phi2_closed}
\phi_2 = e \frac{\Delta_0}{2 \sqrt2\,r_0}  \frac{\Delta\sin\theta}{r^2R^3},
\end{equation}
where 
\be
R(r,\theta) \equiv({\mathfrak r}^2 + {\mathfrak r}_0^2 - 2 {\mathfrak r} {\mathfrak r}_0 \cos\theta -M^2 \sin^2\theta)^{1/2},
\ee
\be
\mfr=r-M \quad \mbox{and} \quad \mfro=r_0-M.
\ee

An entirely analogous computation of $\phi_0$, from Eq.~(\ref{bp}) with spin-weight 1, yields the 
expression
\begin{equation}\label{em_phi0_closed}
\phi_0=- e \frac{\Delta_0}{\sqrt2\ r_0}\frac{\sin\theta}{R^3}.
\end{equation} 
Our aim, however, is to parallel the Cohen-Kegeles treatment of gravitational perturbations, in which the 
metric and the remaining components of the Weyl tensor are constructed from a Hertz potential $\Psi$.
Here the vector potential $A_\alpha$ plays the role of the metric, and $A_\alpha$, $\phi_0$ and $\phi_1$ 
are constructed from an analogous Cohen-Kegeles Hertz potential $\Phi$.

It is worth pointing out here that, although the electromagnetic field is smooth ($C^\infty$) outside 
the particle, its components $\phi_0$ and $\phi_2$ are not smooth scalars.  

\subsection{The Cohen-Kegeles formalism for electromagnetism}

We begin with a review of the Cohen-Kegeles \cite{CohenKegeles,KegelesCohen} formalism for electromagnetic 
fields on type D spacetimes.  Underlying the formalism is the following relation between a vector potential 
satisfying the sourcefree Maxwell equation and a scalar $\bar\Phi$ :

\noindent
{\sl Proposition 1}.  On a vacuum type-D spacetime, let $A_\alpha$ be a smooth vector field of the form
\be\label{em_ck_vectorpotential_full}
A_\alpha=-m_\alpha(D+2\epsilon+\rho)\Phi+l_\alpha(\delta+2\beta+\tau)\Phi+c.c., 
\ee
where $\Phi$ satisfies the sourcefree Teukolsky equation for $s=-1$ (for $\phi_2$), 
\be\label{em_ck_scalar_full}
\left[({\bm\Delta}+\gamma-\bar{\gamma}+\bar\mu)(D+2\epsilon+\rho)-(\bar\delta+\alpha+\bar\beta-\bar\tau)(\delta+2\beta+\tau)\right]\Phi=0.
\ee 
Then $A_\alpha$ satisfies the sourcefree Maxwell equations,  and the complex components of the corresponding field tensor, 
$F_{\alpha\beta}\equiv\nabla_\beta A_\alpha - \nabla_\alpha A_\beta$, are given by 
\bsube\begin{eqnarray}
\phi_0&=&-(D-\epsilon+\bar{\epsilon}-\bar{\rho})(D+2\bar{\epsilon}+\bar{\rho})\bar\Phi,
\label{em_ck_phi0_full}\\
2\phi_1&=&-[(D+\epsilon+\bar{\epsilon}+\rho-\bar{\rho})(\bar{\delta}+2\bar{\beta}+\bar{\tau})
 \nonumber \\
& &+(\bar{\delta}-\alpha+\bar{\beta}-\pi-\bar{\tau})(D+2\bar{\epsilon}+\bar{\rho})]\bar\Phi,
\label{em_ck_phi1_full}\\
\phi_2&=&-\left[(\bar{\delta}+\alpha+\bar{\beta}-\bar{\tau})(\bar{\delta}+2\bar{\beta}+\bar{\tau})-\lambda(D+2\bar{\epsilon}+\bar{\rho})\right]\bar\Phi.
\label{em_ck_phi2_full}
\end{eqnarray}\esube

The vector potential given here is in what Chrzanowski and several subsequent 
authors call the {\em ingoing radiation gauge} (IRG), 
in which $A_\alpha l^\alpha = 0$.  The terminology, however, is misleading: 
An {\em outgoing} radiation field in a Lorenz gauge has, near future 
null infinity,  wave-vector $l^\alpha$ orthogonal to $A_\alpha$ 
(Chrzanowski incorrectly claims in Sect. IV of \cite{chrz}
that this is true of an ingoing field near past null infinity). 

An identical proposition holds for the corresponding ``ORG'' gauge, in which the roles of the outgoing and ingoing null vectors, $l^\alpha$ and $n^\alpha$, are exchanged. 
Because of the incorrect identification of ``outgoing'' and ``ingoing'' in the 
literature we will simply use the term {\em radiation gauges}.   

Two comments on what the proposition does {\em not} imply: Note first that the component $\phi_2$ of the 
electromagnetic field constructed from $A_\alpha$ is not the function $\Phi$, although both $\Phi$ and 
$\phi_0$ satisfy the same sourcefree Teukolsky equation. Second, the fact that $\phi_2$, say, satisfies the sourcefree Teukolsky equation does not guarantee that $\Phi$ satisfies the sourcefree Teukolsky equation.  
Freedom arising from constants of integration must be used to obtain a $\Phi$ that does so.

The expressions for $A_\alpha$ and for the components of the field tensor in terms of $\Phi$ are much 
simpler in a Schwarzschild geometry.  In this case, all spin coefficients are real and most vanish:
\be
 \epsilon=\tau=\lambda=\pi=\alpha+\beta= 0.
\ee
Eqs.~(\ref{em_ck_vectorpotential_full}-\ref{em_ck_phi2_full}) then become
\begin{eqnarray}\label{em_ck_vectorpotential}
A_\alpha&=&- m_\alpha(D+\rho)\Phi+l_\alpha(\delta+2\beta)\Phi+c.c.,
\label{a_scw_irg}\\
\left[ ({\bm\Delta}+\mu)(D+\rho)\right.&-&\left.\bar{\delta}(\delta+2\beta)\right]\Phi=0,
\label{em_ck_scalar}
\eea
\bsube\bea
\phi_0&=&-D^2\bar\Phi\label{em_ck_phi0}, \\
\phi_1&=&-D(\bar{\delta}+2\beta)\bar\Phi,\label{em_ck_phi1}\\
\phi_2&=&-\bar{\delta}(\bar{\delta}+2\beta)\bar\Phi.\label{em_ck_phi2}
\end{eqnarray}\esube

For future reference, we note that each operator involving $\delta$ or $\bar\delta$ is equal to 
$\eth$ or $\bar\eth$ up to factors of $- r\sqrt2$. In particular, Eq.~(\ref{em_ck_phi2}) 
for $\phi_2$ can be written in the form, 
\be
\phi_2=-\frac{1}{2r^2}\bar{\eth}^2\bar\Phi.
\label{phi2}
\ee

\subsection{The scalar potential $\bar\Phi$}

We now invert the CK relation (\ref{phi2}) to find $\bar\Phi$ in terms of the field component $\phi_2$ 
of a static charge, using the closed-form expression (\ref{em_phi2_closed}) for $\phi_2$. 
From Eq.~(\ref{green_eth_bar_a}) and the axisymmetry of $\phi_2$, we see that the leftmost $\bar\eth$ is 
just $-\partial_\theta$. Integrating with respect to $\theta$, we have
\begin{eqnarray}
\bar{\eth}\bar\Phi&=&\frac{ e\ \Delta_0\ \Delta}{\sqrt2\ r_0} 
 \int \left[ \frac{\sin\theta}{R^3(r,\theta)} \right]
    d\theta
\nonumber\\
&=&\frac{e}{\sqrt{2}\ r_0}
   \left[\frac{M^2\cos\theta-\mfr\mfro}{R}+a(r)\right],
\label{em_eth_phi}
\end{eqnarray}
with $a(r)$ an arbitrary function of $r$. 
Similarly, after writing $\bar{\eth} \bar\Phi=-(\sin\theta)^{-1}\partial_\theta(\sin\theta \bar\Phi)$, a second integration 
with respect to $\theta$ yields
\begin{eqnarray}
\bar\Phi&=&\frac{e}{\sqrt2\ r_0} \frac{1}{\sin\theta}
  \int\left[\frac{M^2\cos\theta-\mfr\mfro}R + a(r)\right]d\cos\theta
\nonumber\\
&=&\frac{e}{\sqrt2\ r_0\sin\theta}\left[R+a(r)\cos\theta+b(r)\right],
\label{em_psi_full}\end{eqnarray}
with $b(r)$ again arbitrary.

Eq.~(\ref{em_psi_full}) has the form of a particular + a homogeneous solution to Eq.~(\ref{phi2}),
\begin{eqnarray}
\bar\Phi & =  & \bar\Phi_P+ \bar\Phi_H, 
\label{phiph}\\
\bar\Phi_P&=& \frac{ e}{\sqrt2\ r_0}\frac R{\sin\theta}  \label{em_psi_p}, 
\label{phip}\\
\bar\Phi_H&=& \frac{ e}{\sqrt2\ r_0} \frac{a(r)\cos\theta+b(r)}{\sin\theta}.
\label{phih}
\end{eqnarray}
The particular solution $\bar\Phi_P$ already satisfies the sourcefree Teukolsky equation. The homogeneous solution,
$\bar\Phi_H$, however, does so if and only if $a''(r)=b''(r)=0$. With this restriction, the  
solutions $\bar\Phi_H$ constitute a 4-parameter set, specified by
\be 
a(r)=a_0 + a_1 r, \qquad b(r)=b_0 + b_1 r.
\label{ab}\ee
By construction, $a$ and $b$ encode no information about $\phi_2$.  Because we expect $\phi_2$ to carry 
all information about the $l\geq 1$ part of the field, the value of $\phi_0$ should similarly be 
independent of $a$ and $b$. This follows directly from Eqs.~(\ref{em_ck_phi0}) and (\ref{ab}), which 
together imply that $\phi_0[\bar\Phi_H] = 0$. 
Then $a$ and $b$ can carry information only about the monopole field. As we show in the next section, 
the parameter $a_0$ corresponds to adding charge to the black hole. The remaining parameters $a_1, b_0$ and $b_1$ correspond simply to gauge transformations (with $b_1$ the trivial gauge transformation, altering neither the field nor the vector potential).  

The potential $\bar\Phi$ is in general singular on the $z$ axis because of an overall factor of $1/\sin\theta$.  This singularity is carried to the vector potential as well, because the angular pieces of equation (\ref{em_ck_vectorpotential}) contain only radial derivatives.  We will see that one can choose $a_1$ and $b_0$ to make $A_\alpha$ smooth on the axis 
for $r>r_0$ or for $r<r_0$, but not both.

\subsection{Completion of the solution}

\noindent
{\em Fields corresponding to $\bar\Phi_H$ }\\

The vector potential $A_\alpha[\bar\Phi_H]$, associated with the homogeneous part of the potential\\
\be
 \bar\Phi_H =\frac{ e}{\sqrt2\ r_0\sin\theta} \left[(a_0 + a_1 r)\cos\theta+(b_0 + b_1 r)\right],
\label{phih1}\ee
has in the radiation gauge the value 
\be
A_\alpha[\bar\Phi_H]
= \frac e {2 r_0 } \left[ 
	-a_0\left( \frac1r l_\alpha-\sqrt{2}\,\frac{\cot\theta}r\, m_\alpha\right) 
	-a_1 l_\alpha+b_0\frac{\sqrt2}{r\sin\theta}m_\alpha
\right]+c.c.
\label{AH}
\ee
Using the relations 
\be
l_\alpha = \nabla_\alpha u, \qquad \frac1{r\sin\theta}m_\alpha = -\frac1{\sqrt2}\nabla_\alpha
\left(\log\tan\frac\theta2 +i\phi\right),
\ee
with $u$ the null coordinate \dis u = t-r-2M\ln(r/2M-1)$, we have 
\be
A_\alpha[\bar\Phi_H]
= -a_0 \frac {e}{2 r_0 r} \left(l_\alpha-
\sqrt 2\, \cot\theta\,m_\alpha\right)+c.c. +\nabla_\alpha \Lambda,
\label{ahlambda}\ee
where the gauge scalar $\Lambda$ is given by 
\be
\Lambda = \frac e{2r_0}\left[a_1u - b_0\left(\ln\tan\frac\theta2+i\phi\right)\right]+ c.c.
\ee
Thus, as claimed, the vector potentials associated with $a_1$ and $b_0$ are pure gauge, and that associated with $b_1$ vanishes.  

The identification of $a_0$ with a change in the black hole's charge (electric and magnetic) follows from the form of the 
corresponding field tensor:  By a test charge $q$ on a Schwarzschild black hole is meant a radial electric field,
with \dis F_{\alpha\beta}\hat t^\alpha \hat r^\beta = \frac q{r^2}$.  Equivalently, only the spin-0 part of the field is 
nonzero, and it is real, with value \dis\phi_1=-\frac q{2r^2}$.  From Eqs.~(\ref{em_ck_phi0}-\ref{em_ck_phi2}), the field associated with $\bar\Phi_H$ for $a_0$ real is the field of a black-hole test charge of magnitude \dis Q=\frac{ea_0}{r_0}$:    
\bsube
\bea
\phi_0[\bar\Phi_H] &=& \phi_2[\bar\Phi_H] = 0,\\
\phi_1[\bar\Phi_H] &=& -\frac{ea_0}{2r_0}\frac1{r^2}.
\eea
\esube
Similarly, a test magnetic charge $g$ on a Schwarzschild black hole has the meaning of a radial magnetic field with
\dis F_{\alpha\beta}\hat \theta^\alpha \hat \phi^\beta= \frac g{r^2}$. Equivalently, only $\phi_1$ is nonzero, 
and it is imaginary, with value \dis\phi_1 = \frac g{2r^2}$.  And the field associated with $\bar\Phi_H$ for $a_0$ 
imaginary is the field of a test magnetic charge of magnitude \dis g = \frac{e (\mbox{Im}\,a_0)}{r_0}$.

If we replace $a_0$ in $\bar\Phi_H$ by $Q\equiv r_0 a_0/e$, we can summarize the last paragraph as follows:\\
The electromagnetic field associated with \dis \bar\Phi = \frac Q{\sqrt2}\cot\theta$ is a monopole field with electric charge
Re $Q$ and magnetic charge Im $(-Q)$. \\
 \\ 

\noindent
{\em Fields corresponding to $\bar\Phi_P$ }

The vector potential corresponding to $\bar\Phi_P$ has the value  
\be 
 A_\alpha[\bar\Phi_P]=-\frac{e}{\sqrt{2}r_0
\sin\theta}\left(\frac{{\mathfrak r}-{\mathfrak r}_0\cos\theta}{R}-\frac{R}{r}\right)(m_\alpha+{\overline m}_\alpha)
+\frac{e}{r_0 r R}\left({\mathfrak r}{\mathfrak r}_0-M^2\cos\theta\right)l_\alpha. 
\label{AP}
\ee

The corresponding components of the field, given by Eqs.~(\ref{em_ck_phi0}-\ref{em_ck_phi2}), are 
\bsube
\begin{eqnarray}
\phi_0&=&-e  \frac{\Delta_0}{\sqrt2\ r_0}\frac{\sin\theta}{R^3}
\label{phi0P}, \\
\phi_1[\bar\Phi_P]&=&\partial_r\left(\frac{e}{2 r_0 r}\frac{M^2\cos\theta-\mfr \mfro} R\right),
\label{phi1P}\\
\phi_2&=&e \frac{\Delta_0}{2\sqrt{2}\,r_0}\frac{\Delta\sin\theta}{r^2R^3}.
\label{phi2P}
\end{eqnarray}\esube

As we now show, the black-hole charge for this field is nonzero.  To find the field of a point charge for 
which the black hole has zero charge, one must add to $\phi_1[\bar\Phi_P]$ a multiple of $\phi_1[\bar\Phi_H]$ 
that has black-hole charge opposite to that of our particular solution.  The total charge of the particular solution 
(black-hole charge + point charge) 
can be found by evaluating $4\pi Q=\int_S\vec{E}\cdot d\vec{S}=\int_S 2\phi_1 dS $ over any two-sphere enclosing the black hole and particle.  We can, for simplicity, take the sphere to be at spatial infinity, writing 
\begin{eqnarray}
Q
&=&\frac1{4\pi }\lim_{r\to\infty}\int_{S_r}\partial_r\left(\frac{e}{2 r_0 r}\frac{M^2\cos\theta-\mfr \mfro} R\right)r^2d\Omega\nonumber\\
&=& e \left(1-\frac{M}{r_0}\right).
\end{eqnarray}
Thus the particular linearized solution associated with $\bar\Phi_P$ describes a particle 
of charge $e$ outside a black hole of charge   
\be 
 Q_{\rm BH} = -e\frac M{r_0}.
\label{qbh}\ee 

We will see in the next section that any homogeneous solution $\bar\Phi_H$ that makes the 
vector potential smooth for $r<r_0$ has a black hole charge that exactly cancels the black hole 
charge of the particular solution:  A radiation gauge is smooth for $2M<r<r_0$ only if spheres with 
$2M<r<r_0$ have no enclosed charge.  Similarly, we will see that a radiation gauge is smooth 
for $r>r_0$ only if the total charge (black hole + particle) vanishes:  Spheres with  
$r>r_0$ enclose no charge.

\noindent{\em Parameter choices for solutions smooth on the axis of symmetry}

As we showed in Appendix~\ref{appendixa}, a {\em sufficient} condition 
for $A_\alpha$ to 
be smooth is that the Hertz potential $\bar\Phi$ is smooth; and $\bar\Phi$ of the 
form given by Eqs. (\ref{phiph}) -- (\ref{ab}) is smooth if and 
only if $\bar\Phi = O(\theta)$ near $\theta=0$ and $\bar\Phi = O(\pi-\theta)$ near $\theta=\pi$.
Smoothness of $\bar\Phi$, however, is not a {\em necessary} condition for smoothness of $A_\alpha$, 
because $\bar\Phi$ involves a parameter $b_1$ that corresponds to a vanishing vector potential.  
By omitting conditions on $b_1$, we obtain necessary and sufficient conditions for 
smoothness of $A_\alpha$.  (It is slightly more efficient to find the conditions from $\bar\Phi$, 
but we will also obtain them directly from the components of $A_\alpha$ as a check.)

From Eqs.~(\ref{phip}) and (\ref{phih1}), we have 
\be
\bar\Phi = \bar\Phi_P+ \bar\Phi_H =\frac e{\sqrt2\,r_0\sin\theta}
	[R+(a_0+a_1r)\cos\theta + b_0+b_1r].
\label{Phi}\ee
Smoothness on the $\theta=0$ part of the axis (the part of the axis that 
contains the particle) is enforced by different conditions below and above the particle's position 
at $r=r_0$.  Formally, the difference arises from the fact that the quantity  
\dis R=({\mathfrak r}^2 + {\mathfrak r}_0 ^2 - 2 {\mathfrak r} {\mathfrak r}_0 \cos\theta 
	- M^2 \sin^2\theta)^{1/2}$ 
in the numerator of the above expression for $\bar\Phi$ has the value $R(\theta=0)=|r-r_0|$. 
Then $\bar\Phi$ is smooth at $\theta=0$ if and only if 
\dis a_0+a_1r+b_0+b_1r=-|r-r_0|$, or 
\bsube\label{asmooth0}
\begin{eqnarray}
a_0+b_0&=&-r_0, \qquad a_1+b_1 = 1,\quad r<r_0\\
a_0+b_0&=&r_0, \qquad a_1+b_1 = -1,\quad r> r_0.
\end{eqnarray}\esube
Similarly, from the value \dis R(\theta=\pi)=r+r_0-2M$, it follows that $\bar\Phi$ is smooth at $\theta=\pi$ if and only if 
\be
 a_0-b_0=r_0-2M.
\label{asmoothpi}\ee
The conditions on $a_0$ and $b_0$ in Eqs.~(\ref{asmooth0}) are inconsistent, implying that 
no choice of parameters yields a radiation gauge smooth both below and above the particle.
One can at most choose parameters that satisfy condition~(\ref{asmoothpi}) and one of 
the two conditions (\ref{asmooth0}).  These choices yield a vector potential $A_\alpha$  \\
(I) smooth everywhere except $\theta=0, \ r>r_0$, or\\
(II) smooth everywhere except $\theta=0, \ r<r_0$.\\
The implied choice of parameters for each case is given in Table~\ref{em_regularity}.
\begin{table}[h]
\caption{Singularity location for two choices of parameters}
\label{em_regularity}
\begin{center}
\begin{tabular}{| c | c | c |}
\hline
Parameter  & Singularity at &Singularity at\\
 & $\theta=0$; $r\geq r_0$ & $\theta=0$; $r\leq r_0$\\
\hline
$a_0$ & $-M$ & $r_0-M$\\
$b_0$ & $-(r_0-M)$ & $M$ \\
\hline
\end{tabular}
\end{center}
\end{table}
Note that any choice of the parameter $a_1$ is permitted, because one can choose 
the trivial parameter $b_1$ to satisfy the second condition 
without altering the value $A_\alpha$.  

In case (I) the vector potential is, in particular, smooth in the exterior 
Schwarzschild geometry for all $r<r_0$.  The parameter value $a_0=M$ corresponds to 
a black hole charge $eM/r_0$ that exactly cancels the black hole charge (\ref{qbh}) 
of the particular solution.  In case (II) the vector potential is smooth for all $r>r_0$, 
and the parameter value $a_0=r_0-M$ corresponds to a black hole charge $-e(1-M/r_0)$ 
that cancels the total charge of the particular solution: The full solution has 
a black hole with charge $-e$ and total charge 0. Thus as noted above, a Hertz potential 
$\bar\Phi$ yields a smooth radiation gauge on a sphere only if the total electric and magnetic flux 
through the sphere vanish. 

This is not surprising.  The radiation gauges are designed for sourcefree solutions and 
are constructed from the spin $\pm 1$ parts of the field - in our case from parts of the 
field with no $l=0$ part.  For a point-charge with no charge on the black hole, we can find regular 
radiation gauges outside and inside the charge for the $l \ge 1$ part of the field and can add to them the vector potential in another gauge for the Coulomb $l=0$ field.

It makes sense to use a radiation gauge to describe the part of the field with spin greater 
than 0 and to add an $l=0$ part in another gauge.  In particular, by 
adding the solution describing a perturbation of the black hole's 
charge in a Coulomb gauge, namely 
\bea
A_t &=& \frac Qr, \nonumber \\
A_\alpha &=& A_t \nabla_\alpha t = \frac Q{r}\left(\frac12 l_\alpha + \frac{r^2}\Delta n_\alpha\right),
\label{coulomb}\eea
we obtain a solution with arbitrary black hole charge, whose gauge singularity is 
restricted to lie on the $\theta=0$ ray, either above or below the particle. 
\footnote{
Although one cannot obtain the $l=0$ vector potential in a radiation gauge 
by using a Hertz potential of the form (\ref{Phi}), the vector potential
{\em can} be expressed in a radiation gauge:  
Writing \dis f=Q\ln\left(\frac r{2M} -1\right)$, we have 
\dis\tilde A_\alpha\equiv A_\alpha -\nabla_\alpha f = \frac Qr l_\alpha$, 
implying $\tilde A_\alpha l^\alpha = 0$.
}

The vector potentials corresponding to cases 
(I) (singular only for $\theta=0, r\geq r_0$) and \\
(II) (singular only for $\theta=0, r\leq r_0$) are, respectively, 
\bsube
\begin{eqnarray}
A_\alpha^{(I)}&=&
	\frac e{r_0r}\left(\frac{{\mathfrak r}{\mathfrak r}_0-M^2\cos\theta}R+M\right)l_\alpha
\nonumber\\
 &&+\frac e{\sqrt2\,r_0\sin\theta}
   \left(\frac{R-{\mathfrak r}_0-M\cos\theta}r-\frac{{\mathfrak r}-{\mathfrak r}_0\cos\theta}R\right)
   (m_\alpha+\overline m_\alpha),
\label{a1}\\
A_\alpha^{(II)}&=&
	\frac e{r_0r}\left(\frac{{\mathfrak r}{\mathfrak r}_0-M^2\cos\theta}R-{\mathfrak r}_0\right)l_\alpha
\nonumber\\
 &&+\frac e{\sqrt2\,r_0\sin\theta}
   \left(\frac{R+M+{\mathfrak r}_0\cos\theta}r-\frac{{\mathfrak r}-{\mathfrak r}_0\cos\theta}R\right)
   (m_\alpha+\overline m_\alpha).
\label{a2e}
\end{eqnarray}\esube
Here we have set the parameter $a_1$ to zero.  

As we have noted, Eqs.~(\ref{a1}) and (\ref{a2e}) are physically different solutions, with the 
former having an uncharged black hole, the latter a black hole with charge $-e$.  
Adding to the vector potential (\ref{a2e}) the $l=0$ (Coulomb) perturbation (\ref{coulomb}) that 
cancels the charge \dis -e$ on the black hole, we obtain 
a vector potential that is gauge related to that of Eq.~(\ref{a1}) and is smooth except 
on the axis below the particle:
\bea 
A_\alpha^{(II)} &=&
	\frac e{r_0r}\left(\frac{{\mathfrak r}{\mathfrak r}_0-M^2\cos\theta}R-\frac 1 2\frac{\Delta_0}{r_0}\right)l_\alpha
	+ \frac{er}\Delta n_\alpha
\nonumber\\
 &&+\frac e{\sqrt2\,r_0\sin\theta}
   \left(\frac{R+M+{\mathfrak r}_0\cos\theta}r-\frac{{\mathfrak r}-{\mathfrak r}_0\cos\theta}R\right)
   (m_\alpha+\overline m_\alpha).
\label{a2}\eea

Finally, as a check, we directly verify the smoothness conditions governing $a_0$ and $b_0$ 
in (\ref{asmooth0}) and (\ref{asmoothpi}), using the components of 
$A_\alpha[\bar\Phi]= A_\alpha[\bar\Phi_P]+A_\alpha[\bar\Phi_H]$ 
given by Eqs.~(\ref{AH}) and (\ref{AP}). Because $A_2$ is already smooth for arbitrary values 
of the parameters, $A_\alpha$ will be smooth if and only if $A_4$ is $O(\theta)$ near $\theta=0$ 
and $O(\pi-\theta)$ near $\theta=\pi$.  Near $\theta=0$, we have 
\bsube
\begin{eqnarray}
A_4[\bar\Phi]&=&-\frac{e}{\sqrt{2}r_0r}\frac{a_0+b_0+r_0}\theta+O(\theta),\quad r<r_0\\
A_4[\bar\Phi]&=&-\frac{e}{\sqrt{2}r_0r}\frac{a_0+b_0-r_0}\theta+O(\theta),\quad r> r_0.
\end{eqnarray}\esube

Near the axis of symmetry on the side of the black hole opposite to the particle ($\theta=\pi$) the corresponding expansion of $A_\alpha$ has the form 
\be
A_4=-\frac{e}{\sqrt{2}\,r_0r}\frac{-a_0+b_0+r_0-2M}{\pi-\theta}+O[(\pi-\theta)].
\ee
The conditions on $a_0$ and $b_0$ in Eqs.~(\ref{asmooth0}) and (\ref{asmoothpi}) are exactly 
conditions that the coefficients of $\theta^{-1}$ and $(\pi-\theta)^{-1}$ vanish.   

We can explicitly verify that the vector potentials $A^{(I)}_\alpha$ and $A^{(II)}_\alpha$ we 
have obtained for a point charge outside an uncharged black hole are related by a gauge 
transformation to the solution obtained by Copson.  In particular, using the gauge function 
\begin{eqnarray}\label{em_gauge_transform}
\Lambda(t,r,\theta,\phi)&=&-e \ln \left[ \frac{{\mathfrak r}_0 \cos\theta-{\mathfrak r}+R}{\cos\theta+1}\right] + \frac{e M}{r_0} \ln \left[ 2 \left( {\mathfrak r}_0 \cos\theta- {\mathfrak r}+R \right) \right]
\nonumber\\
&&+\frac{e M}{r_0} \ln \left[\frac{\Delta_0-M^2\sin^2\theta + M r-{\mathfrak r}_0 r \cos\theta+\left(M(1+\cos\theta)-r_0 \right)R}{M(1+\cos\theta)\sin^2\theta} \right],
\end{eqnarray}
we find $A_\alpha^{(I)}+\nabla_\alpha \Lambda = A_t \nabla_\alpha t$, with 
\begin{eqnarray}\label{em_at}
A_t&=&\frac{e}{r r_0}\left(\frac{{\mathfrak r}{\mathfrak r}_0-M^2\cos\theta}{R}+M\right).
\end{eqnarray}

\section{Static Gravitational Perturbations}
\label{mass1}

\subsection{Outline}

\noindent {\sl Notation}. 
Because the components $\psi_0, \psi_1, \psi_3$ and $\psi_4$ of the Weyl tensor vanish in the 
background Schwarzschild spacetime, we use the same symbols to denote the components of the 
perturbed Weyl tensor.  Because $\Psi_2$ is nonzero in the background spacetime, we 
denote the perturbed component by $\psi_2$.  \\

\noindent
{\sl Outline of Computation}.
To compute the linearized gravitational field of a point mass in the Cohen-Kegeles formalism, we roughly 
parallel the calculation of the electromagnetic field of a point charge, but there are two primary 
differences. First, where the electromagnetic 
field tensor is gauge invariant, only the extreme spin components, $\psi_0$ and $\psi_4$, 
of the perturbed Weyl tensor are gauge invariant. 
Second, the electromagnetic field of a static test charge makes sense, but a static  
test mass must be supported by something that itself contributes to the linearized gravitational 
field, and in our solution this is seen as a conical singularity (a deficit angle) along the axis of symmetry inside or outside $r=r_0$.  Nevertheless, the extreme spin components 
of the perturbed Weyl tensor are smooth everywhere except at the 
position of the particle: The part of the perturbation with $|s|=2$ does not know about the strut.
As in the electromagnetic case, we obtain simple closed-form expressions for these gauge-invariant 
parts of the static perturbation.   

We first use the decoupled equation (\ref{bp}) to find $\psi_4$, again by taking angular derivatives 
of the static Green's function. 
Using a radiation gauge, we construct a potential 
$\Psi$, in terms of which we compute the components $h_{\alpha\beta}$ of the perturbed metric 
and the remaining components $\psi_1,\psi_2$ and $\psi_3$ of the field. In so doing, we correct of factor-of-two 
error in the Cohen-Kegeles formalism.
    
Because $\psi_4$ has spin-weight $-2$, it has no $l=0$ or $l=1$ parts. The monopole part of the field 
corresponds to a change in the mass or area of the black hole, the dipole part to a change in its 
angular momentum and center of mass.  As in the case of the electromagnetic charge, the change in the 
black hole's mass must be specified separately.  The change in its center of mass appears as a gauge 
transformation. The potential $\Psi$ is obtained from $\psi_4$ by solving a fourth-order differential 
equation, and the freedom to add mass and spin to the black hole can be identified with two of the  
the constants of integration. We again find local gauges that are singular on different parts of the 
axis of symmetry (the singularities in these radiation gauges were identified by Barack and Ori\cite{BarackOri}), with a residual conical singularity (the strut) that cannot be removed by a choice of 
gauge. 
   
\subsection{Computing $\psi_4$ from the decoupled wave equation}

Consider a static point particle of mass $\mathfrak m$ at a point $x_0$ with coordinates 
$r=r_0$, $\theta=0$. The particle has density 
\be
\rho_m(x)={\mathfrak m} \delta^3(x,x_0)
\ee
and stress-energy tensor 
\be 
T^{\alpha\beta} = \rho_m u^\alpha u^\beta, \qquad \mbox{with } u^\alpha = \frac r{\Delta^{1/2}}t^\alpha.
\ee
In the source term of the Bardeen-Press equation for  $\rho^{-4}\psi_4$, given in Table \ref{sourcetable}, 
only the component $T_{22}$ is nonzero, and the source is simply  
\be
T=-2(\bar{\delta}-2\beta)\bar{\delta}\rho^{-4}T_{22}, 
\ee
with 
\begin{eqnarray}\label{gr_source}
T_{22}\equiv T_{\alpha\beta}n^\alpha n^\beta={\mathfrak m}\frac{ \Delta^{3/2}}{4r^5}\delta(r-r_0)
\delta(\cos\theta-1)\delta(\phi).
\end{eqnarray}

By writing the angular derivatives in the source in terms of $\eth$ and using equation 
(\ref{green_sylm_prop_5}), we can expand the source in spin-weighted spherical harmonics: 
\begin{eqnarray}
T(x)&=&-2(\bar{\delta}-2\beta)\bar\delta\rho^{-4}T_{2 2}\nonumber\\
&=& - r^2\bar{\eth}^2T_{2 2}\nonumber\\
&=& - {\mathfrak m} \frac{\Delta^{3/2}}{4r^3}\delta(r-r_0)\bar{\eth}^2\sum_{l=0}^\infty\sum_{m=-l}^lY_{l m}(\theta,\phi)Y_{l m}^*(0,0)\nonumber\\
&=& - {\mathfrak m}\frac{ \Delta^{3/2}}{4r^3}\delta(r-r_0)\sum_{l=2}^\infty\sum_{m=-l}^l\left[\frac{(l+2)!}{(l-2)!}\right]^{1/2}\!\!{}_{-2}Y_{l m}(\theta,\phi)Y_{l m}^*(0,0).
\end{eqnarray}
We can now obtain $\psi_4$ from the Green's function of Eq.~(\ref{green_greens_function}).  
Eq.~(\ref{green_definitions_b}) for $\rho^{-4}\psi_4$ has the form,
\begin{eqnarray}\label{gr_psi4_integral}
\rho^{-4}\psi_4&=&\int G(x,x')T(x')\sqrt{-g}d^3x'\\
&=&\int\left[ \frac{ 4 \pi\Delta}{M \Delta'} \sum_{l=2} ^{\infty}\sum_{m=-l} ^l \frac{(l-2)!} {(l+2)!} P_l ^2 \left( \frac{{\mathfrak r}_<}{M} \right) Q_l ^2 \left( \frac{{\mathfrak r}_> }{M}\right)\ _{-2}Y _{l m}(\theta, \phi)\ _{-2}Y^* _{l m} (\theta', \phi')
\right]\nonumber\\
& &\times \left[ - \frac{{\mathfrak m} {\Delta'}^{3/2}}{4{r'}^3}\delta(r'-r_0)\sum_{n=2}^\infty\sum_{p=-n}^n\left[\frac{(n+2)!}{(n-2)!}\right]^{1/2}{}_{-2}Y_{n p}(\theta',\phi')Y_{n p}^*(0,0)\right] r'^2dr'd\Omega',
\eea
and, using orthogonality of the spin-weighted harmonics, we have
\be
\psi_4= - \frac{\pi {\mathfrak m}}{M} \frac{\Delta\Delta_0^{1/2}}{r^4 r_0}\sum_{l=2}^\infty \sum_{m=-l}^l \left[\frac{(l-2)!}{(l+2)!}\right]^{1/2}P_l^2\left(\frac{{\mathfrak r}_<}{M}\right) Q_l^2\left(\frac{{\mathfrak r}_>}{M}\right) {}_{-2}\!Y_{l m}(\theta,\phi) Y_{l m}^*(0,0).
\label{psi4_mode}
\ee
It is again possible to sum this series to obtain the closed-form expression (details in Appendix \ref{appendixb}), 

\be\label{gr_psi4_closed}
\psi_4 = - {\mathfrak m}\frac{3\Delta_0^{3/2}}{4 r_0} 
		\frac{\Delta^2\sin^2\theta } {r^4 R^5}.
\ee
An entirely analogous computation of $\psi_0$, from Eq.~(\ref{bp}) with spin-weight 2, yields the 
expression
\begin{equation}\label{gr_psi0_closed}
\psi_0=- {\mathfrak m} \frac{3\Delta_0^{3/2}}{r_0}\frac{\sin^2\theta}{R^5}.
\end{equation} 

\subsection{The Cohen-Kegeles formalism for gravity}

The formalism of Cohen and Kegeles \cite{KegelesCohen} again relates $\psi_4$ to a scalar potential $\Psi$ which we can then use to reconstruct the perturbed metric $h_{\alpha\beta}$.  In each of the key relations, 
Eqs.~(\ref{kc_psi_0}-\ref{kc_psi_4}) below, we correct an error that appears in 
the Cohen-Kegeles papers and has been repeated in the subsequent literature: The right hand side of 
each equation is reduced by the factor $1/2$ from the corresponding Cohen-Kegeles expressions.    

\noindent
{\sl Proposition 2}.  On a vacuum type-D spacetime, let $h_{\alpha\beta}$ be a smooth tensor field of the form
\begin{eqnarray}\label{kc_metric}
h_{\alpha\beta}&=&-\{\, l_\alpha l_\beta \left[(\bar{\delta}+\alpha+3\bar{\beta}
    -\bar{\tau})(\bar{\delta}+4\bar{\beta}+3\bar{\tau})-\lambda(D+4\bar{\epsilon}+3\bar{\rho})\right]\nonumber\\
& &+{\overline m}_\alpha{\overline m}_\beta (D-\epsilon+3\bar{\epsilon}-\bar{\rho})(D+4\bar{\epsilon}+3\bar{\rho}) \nonumber\\
& &-l_{(\alpha}{\overline m}_{\beta)}[(D+\epsilon+3\bar{\epsilon}+\rho-\bar{\rho})(\bar{\delta}+4\beta+3\bar{\tau})+\nonumber\\
& &(\bar{\delta}-\alpha+3\bar{\beta}-\pi-\bar{\tau})(D+4\bar{\epsilon}+3\bar{\rho}) ]\, \} \bar{\Psi}+c.c.,
\end{eqnarray}
where $\Psi$ satisfies the sourcefree Teukolsky equation for $s=-2$ (for $\psi_4$), 
\begin{eqnarray}\label{kc_scalar_wave}
\left[(\delta+3\bar{\alpha}+\beta-\tau)(\bar{\delta}+4\bar{\beta}+3\bar{\tau})-({\bm\Delta}-\gamma+3\bar{\gamma}+\mu)(D+4\bar{\epsilon}+3\bar{\rho})+3\bar{\Psi}_2\right]\bar{\Psi}=0.
\end{eqnarray}
Then $h_{\alpha\beta}$ satisfies the perturbed vacuum Einstein equation, and the complex components of the corresponding perturbed Weyl tensor are given by 

\begin{eqnarray}
\psi_0&=&\frac{1}{2}(D-3\epsilon+\bar{\epsilon}-\bar{\rho})(D-2\epsilon+2\bar{\epsilon}-\bar{\rho})(D-\epsilon+3\bar{\epsilon}-\bar{\rho})(D+4\bar{\epsilon}+3\bar{\rho})\bar{\Psi},\label{kc_psi_0}\\
\psi_1&=&\frac{1}{8}[(D-\epsilon+\bar{\epsilon}+\rho-\bar{\rho})(D+2\bar{\epsilon}+\rho-\bar{\rho})(D+\epsilon+3\bar{\epsilon}+\rho-\bar{\rho})(\bar{\delta}+4\bar{\beta}+3\bar{\tau})\nonumber\\
& &+(D-\epsilon+\bar{\epsilon}+\rho-\bar{\rho})(D+2\bar{\epsilon}+\rho-\bar{\rho})(\bar{\delta}-\alpha+3\bar{\beta}-\pi-\bar{\tau})(D+4\bar\epsilon+3\bar{\rho})\nonumber\\
& &+(D-\epsilon+\bar{\epsilon}+\rho-\bar{\rho})(\bar{\delta}-2\alpha+2\bar{\beta}-2\pi-\bar{\tau})(D-\epsilon+3\bar{\epsilon}-\bar{\rho})(D+4\bar\epsilon+3\bar{\rho})\nonumber\\
& &+(\bar{\delta}-3\alpha+\bar{\beta}-3\pi-\bar{\tau})(D-2\epsilon+2\bar{\epsilon}-\bar{\rho})(D+\epsilon+3\bar{\epsilon}-\bar{\rho})(D+4\bar{\epsilon}+3\bar{\rho})]\bar{\Psi},\label{kc_psi_1}\\
\psi_2&=&\frac{1}{12}[(D+\epsilon+\bar{\epsilon}+2\rho-\bar{\rho})(D+2\epsilon+2\bar{\epsilon}+2\rho-\bar{\rho})(\bar{\delta}+\alpha+3\bar{\beta}-\bar{\tau})(\bar{\delta}+4\bar{\beta}+3\bar{\tau})\nonumber\\
& &+(D+\epsilon+\bar{\epsilon}+2\rho-\bar{\rho})(\bar{\delta}+2\bar{\beta}-\pi-\bar{\tau})(D+\epsilon+3\bar{\epsilon}+\rho-\bar{\rho})(\bar{\delta}+4\bar{\beta}+3\bar{\tau})\nonumber\\
& &+(D+\epsilon+\bar{\epsilon}+2\rho-\bar{\rho})(\bar{\delta}+2\bar{\beta}-\pi-\bar{\tau})(\bar{\delta}-\alpha+3\bar{\beta}-\pi-\bar{\tau})(D+4\bar{\epsilon}+3\bar{\rho})\nonumber\\
& &+(\bar{\delta}-\alpha+\bar{\beta}-2\pi-\bar{\tau})(\bar{\delta}-2\alpha+2\bar{\beta}-2\pi-\bar{\tau})(D-\epsilon+3\bar{\epsilon}-\bar{\rho})(D+4\bar{\epsilon}+3\bar{\rho})\nonumber\\
& &+(\bar{\delta}-\alpha+\bar{\beta}-2\pi-\bar{\tau})(D+2\bar{\epsilon}+\rho-\bar{\rho})(\bar{\delta}-\alpha+3\bar{\beta}-\pi-\bar{\tau})(D+4\bar{\epsilon}+3\bar{\rho})\nonumber\\
& &+(\bar{\delta}-\alpha+\bar{\beta}-2\pi-\bar{\tau})(D+2\bar{\epsilon}+\rho-\bar{\rho})(D+\epsilon+3\bar{\epsilon}+\rho-\bar{\rho})(\bar{\delta}+4\bar{\beta}+3\bar{\tau})]\bar{\Psi},\label{kc_psi_2}\\
\psi_3&=&\frac{1}{8}[(D+3\epsilon+\bar{\epsilon}+3\rho-\bar{\rho})(\bar{\delta}+2\alpha+2\bar{\beta}-\bar{\tau})(\bar{\delta}+\alpha+3\bar{\beta}-\bar{\tau})(\bar{\delta}+4\bar{\beta}+3\bar{\tau})\nonumber\\
& &+(\bar{\delta}+\alpha+\bar{\beta}-\pi-\bar{\tau})(D+2\epsilon+2\bar{\epsilon}+2\rho-\bar{\rho})(\bar{\delta}+\alpha+3\bar{\beta}-\bar{\tau})(\bar{\delta}+4\bar{\beta}+3\bar{\tau})\nonumber\\
& &+(\bar{\delta}+\alpha+\bar{\beta}-\pi-\bar{\tau})(\bar{\delta}+2\bar{\beta}-\pi-\bar{\tau})(D+\epsilon+3\bar{\epsilon}+\rho-\bar{\rho})(\bar{\delta}+4\bar{\beta}+3\bar{\tau})\nonumber\\
& &+(\bar{\delta}+\alpha+\bar{\beta}-\pi-\bar{\tau})(\bar{\delta}+2\bar{\beta}-\pi-\bar{\tau})(\bar{\delta}-\alpha+3\bar{\beta}-\pi-\bar{\tau})(D+4\bar{\epsilon}+3\bar{\rho})]\bar{\Psi},\label{kc_psi_3}\\
\psi_4&=&\frac{1}{2}\{(\bar{\delta}+3\alpha+\bar{\beta}-\bar{\tau})(\bar{\delta}+2\alpha+2\bar{\beta}-\bar{\tau})(\bar{\delta}+\alpha+3\bar{\beta}-\bar{\tau})(\bar{\delta}+4\bar{\beta}+3\bar{\tau})\bar{\Psi}\nonumber\\
 & &+3\Psi_2[\tau(\bar{\delta}+4\alpha)-\rho({\bm\Delta}+4\gamma)-\mu(D+4\epsilon)+\pi(\delta+4\beta)+2\Psi_2]\Psi\}.\label{kc_psi_4}
\end{eqnarray}

Note that the metric $h_{\alpha\beta}$ is tracefree and satisfies $h_{\alpha\beta}l^\beta = 0$. Because these 
are five real conditions on the components of $h_{\alpha\beta}$, and one has only four gauge degrees of freedom, 
one cannot find a radiation gauge for a generic metric.  As 
Price, Shankar and Whiting show, however \cite{WhitingPrice,PriceShankarWhiting}, such a gauge  exists locally for vacuum perturbations of any type D vacuum spacetime. 
In particular a perturbation that changes the mass of a Kerr spacetime 
(with $J$ and $M$ nonzero for the background) cannot be written in a radiation gauge, 
but radiative perturbations and a perturbation that changes the angular 
momentum of Kerr can be.  

As we 
show below (and as Price et al. found independently), however, one {\em can} express 
a perturbation that changes the mass of a Schwarzschild black hole in a radiation
gauge. We exhibit two examples acquired from a Hertz potential that are each 
singular on a ray. Although the Hertz potential does not yield a form of the 
mass perturbation that is nonsingular in the entire exterior Schwarzschild 
geometry, the family of radiation gauges includes such a perturbation, 
and we present it below.  The Hertz potential formalism, appropriate for spin-weight greater than two, yields at least for $s=0$ perturbations (for $l=0$ perturbations of Schwarzschild), a restricted family of radiation gauges.  

By exchanging the roles of the ingoing and outgoing null vectors one obtains  analogous propositions for a radiation gauge in which 
$h_{\alpha\beta}n^\beta = 0, \ h=0$. 

These equations are much simpler on a Schwarzschild background.  The form of $h_{\alpha\beta}$ is 
\begin{eqnarray}\label{gr_scw_irg}
h_{\alpha\beta} = -[ l_\alpha l_\beta (\bar{\delta} + 2\beta)(\bar{\delta} + 4\beta) + {\overline m}_\alpha {\overline m}_\beta (D - \rho)(D + 3\rho) \nonumber \\
- 2 l_{(\alpha} {\overline m}_{\beta)} (D+\rho) (\bar{\delta} + 4\beta) ] \bar{\Psi} + c.c.,
\end{eqnarray}
where
\begin{eqnarray}\label{gr_scw_scalar_eqn}
\left[({\delta}-2\beta)(\bar\delta+4\beta)-({\bm\Delta}+2\gamma+\mu)(D+3\rho)+3\Psi_2\right]\bar\Psi=0.
\end{eqnarray}
The expressions for the complex components of the Weyl tensor become 
\begin{eqnarray}
\psi_0&=&\frac{1}{2}D^4\bar{\Psi} \label{gr_scw_psi0},\\
\psi_1&=&\frac{1}{2}D^3(\bar{\delta}+4\beta)\bar{\Psi}, \label{gr_scw_psi1}\\
\psi_2&=&\frac{1}{2}D^2(\bar{\delta}+2\beta)(\bar{\delta}+4\beta)\bar{\Psi}, \label{gr_scw_psi2}\\
\psi_3&=&\frac{1}{2}D\bar{\delta}(\bar{\delta}+2\beta)(\bar{\delta}+4\beta)\bar{\Psi}, \label{gr_scw_psi3} \\
\psi_4&=&\frac{1}{2}(\bar{\delta}-2\beta)\bar{\delta}(\bar{\delta}+2\beta)(\bar{\delta}+4\beta)\bar{\Psi}\nonumber \\
 & &-\frac{3}{2}\Psi_2\left[\mu D+\rho(\Delta+4\gamma)-2\Psi_2\right] \Psi. \label{gr_scw_psi4}
\end{eqnarray}

In the next section, we solve Eq.~(\ref{gr_scw_psi4}) to find the scalar potential 
$\Psi$ from the closed-form expression (\ref{gr_psi4_closed}) for $\psi_4$.  In the second line of Eq.~(\ref{gr_scw_psi4}), the expression 
in brackets becomes simply \dis-\frac1{2r}\partial_t$, when one substitutes the Schwarzschild values of the operators
and spin coefficients (\ref{spincoefficients}), and the value $-M/r^3$ of the unperturbed $\Psi_2$. In our static case, 
the second line then vanishes, leaving a set of angular derivatives of 
$\bar{\Psi}$.  Replacing $\psi_4$ on the left side by its value 
(\ref{gr_psi4_closed}), we have

\begin{eqnarray}
- \frac{3 {\mathfrak m} \Delta_0^{3/2}}{4  r_0} 
\frac{\Delta^2 \sin^2\theta}{r^4 R^{5}}
&=&\frac{1}{2}(\bar{\delta}-2\beta)\bar{\delta}(\bar{\delta}+2\beta)(\bar{\delta}+4\beta)\bar{\Psi} \\
&=&\frac{1}{8 r^4}\bar{\eth}^4\bar{\Psi},
\label{gr_psi4_eths}
\end{eqnarray}
or 
\begin{eqnarray}
\bar{\eth}^4\bar{\Psi}&=&- \frac{6 {\mathfrak m} \Delta_0^{3/2}}{r_0} 
\frac{\Delta^2 \sin^2\theta}{R^5}.
\label{ethPsi}
\end{eqnarray}

\subsection{The scalar potential $\Psi$}

We can now begin the integration of Eq.~(\ref{ethPsi}) for $\Psi$.  
Because $\psi_4$ has spin-weight $-2$ and $\bar{\eth}$ 
lowers the spin-weight by 1, $\bar{\Psi}$ 
has spin-weight $+2$.  From the definition (\ref{green_eth_bar_a}) of $\bar\eth$, 
its action on the spin-weight $-1$ quantity $\bar{\eth}^3\bar{\Psi}$ is given 
by 
\begin{eqnarray}
\bar{\eth}^4\bar{\Psi}&=&-\sin\theta \partial_\theta \left(\frac{1}{\sin\theta}\bar{\eth}^3\bar{\Psi}\right).
\end{eqnarray}

Integrating this equation with respect to $\theta$, we have
\begin{eqnarray}
\bar{\eth}^3\bar{\Psi}&=& \frac{6{\mathfrak m}\Delta^2\Delta_0^{3/2}}{r_0}\sin\theta\int\frac{\sin\theta d\theta}{R^5}\\
&=&-\frac{6{\mathfrak m}\Delta^2\Delta_0^{3/2}}{r_0}\sin\theta\int\frac{d\cos\theta}{\left({\mathfrak r}^2 + {\mathfrak r}_0 ^2 - 2 {\mathfrak r}{\mathfrak r}_0 \cos\theta - M^2 \sin^2\theta\right)^{5/2}}\\
&=&\frac{2{\mathfrak m}}{r_0\Delta_0^{1/2}} \sin\theta 
\Big\{ \frac{M^2\cos\theta-{\mfr}{\mfro}}{R^3} 
\left[\mfr^2\mfro^2-3M^2({\mathfrak r}^2+{\mathfrak r}_0^2)
       +3M^4+4{\mathfrak r}{\mathfrak r}_0M^2\cos\theta-2M^4\cos^2\theta\right]   \nonumber \\
& & \phantom{xxxxxxxxx}+ a(r) \Big\},
\end{eqnarray}
with $a(r)$ arbitrary.  

Next, acting on the spin-weight 0 quantity $\eth^2\Psi$, $\eth$ has the form
\be
\bar{\eth}^3\bar{\Psi}=-\partial_\theta \bar{\eth}^2\bar{\Psi},
\ee
and we have 
\begin{eqnarray}
\bar{\eth}^2\bar{\Psi}&=& \frac{2{\mathfrak m}}{r_0\Delta_0^{1/2}} 
\int \Big\{ \frac{{M^2\cos\theta-\mathfrak r}{\mathfrak r}_0 }{R^3} 
\left[{\mathfrak r}^2{\mathfrak r}_0^2-3M^2({\mathfrak r}^2+{\mathfrak r}_0^2)+3M^4+4{\mathfrak r}{\mathfrak r}_0M^2\cos\theta-2M^4\cos^2\theta\right]  \nonumber\\
& & \qquad\qquad\qquad + a(r) \Big\} d\cos\theta\\
&=&\frac{2{\mathfrak m}}{r_0\Delta_0^{1/2}} \left[ -\frac{{\mathfrak r}^2{\mathfrak r}_0^2+M^2({\mathfrak r}^2+{\mathfrak r}_0^2)-M^4-4{\mathfrak r}{\mathfrak r}_0M^2\cos\theta+2M^4\cos^2\theta}{\left({\mathfrak r}^2 + {\mathfrak r}_0 ^2 - 2 {\mathfrak r}{\mathfrak r}_0 \cos\theta - M^2 \sin^2\theta\right)^{1/2}}\right.\nonumber\\
& &\left.\phantom{\frac AB}\qquad\qquad+a(r)\cos\theta+b(r)\right],
\end{eqnarray}
with $b(r)$ arbitrary.  

Continuing in this way, we have
\begin{eqnarray}
\bar{\eth}^2\bar{\Psi}&=&-\frac{1}{\sin\theta}\partial_\theta\left(\sin\theta\bar{\eth}\bar{\Psi}\right),
\end{eqnarray}
whose integration entails a third free function $c(r)$:
\begin{eqnarray}
\sin\theta\bar{\eth}\bar{\Psi}
 &=& \frac{2{\mathfrak m}}{r_0\Delta_0^{1/2}} \int \left[ -\frac{{\mfr}^2{\mfro}^2+M^2({\mfr}^2+{\mfro}^2)-M^4-4{\mfr}{\mfro}M^2\cos\theta+2M^4\cos^2\theta}{R}\right.\nonumber\\
& &\left.\phantom{\frac AB}\qquad\qquad+a(r)\cos\theta+b(r)\right] d\cos\theta
\nonumber\\
&=& \frac{2{\mathfrak m}}{r_0\Delta_0^{1/2}} \left[({\mfr}{\mfr}_0-M^2\cos\theta)R
    +\frac{1}{2}a(r)\cos^2\theta+b(r)\cos\theta+c(r)\right].
\end{eqnarray}

The final integration of 
\begin{equation}
\sin\theta\bar{\eth}\bar{\Psi}=-\frac{1}{\sin\theta}\partial_\theta\left(\sin^2\theta\bar{\Psi}\right),
\end{equation}
yields 
\begin{eqnarray}
\sin^2\theta\bar{\Psi}&=&\frac{2{\mathfrak m}}{r_0\Delta_0^{1/2}} \int \left[({\mathfrak r}{\mathfrak r}_0-M^2\cos\theta)R+\frac{1}{2}a(r)\cos^2\theta+b(r)\cos\theta+c(r)\right] d\cos\theta
\nonumber\\
&=&\frac{2{\mathfrak m}}{r_0\Delta_0^{1/2}}\left[ - \frac13 R^3
+\frac{1}{6}a(r)\cos^3\theta+\frac{1}{2}b(r)\cos^2\theta+c(r)\cos\theta+d(r)\right].
\end{eqnarray}
We thus find a remarkably simple particular solution $\bar\Psi_P$, together 
with a homogeneous solution 
$\psi_H$ involving 
four free functions, $a(r), b(r), c(r)$ and $d(r)$: 
\begin{eqnarray}\label{potential}
\bar{\Psi}&=&\bar{\Psi}_P+\bar{\Psi}_H\\
\bar{\Psi}_P&=& - \frac23 A\frac{R^3}{\sin^2\theta}
\label{PsiP}\\
\bar{\Psi}_H&=&2A\frac{1}{\sin^2\theta}\left[\frac{1}{6}a(r)\cos^3\theta+\frac{1}{2}b(r)\cos^2\theta+c(r)\cos\theta+d(r)\right],
\label{PsiH}
\end{eqnarray}
where we denote by $A$ a constant that appears repeatedly throughout this section,
\be 
 A\equiv\frac{\mathfrak m}{r_0\Delta_0^{1/2}}.
\ee

Again the particular solution $\bar\Psi_P$ already satisfies the homogeneous Bardeen-Press equation for $s=2$. Requiring that $\Psi_H$ also be a solution 
restricts each free function to a polynomial whose form is given in 
Table \ref{gr_consts}.  Then the scalars $\bar\Psi$ of Eq.~(\ref{potential}) form 
an eight-complex-parameter set of potentials that generate vacuum perturbations for 
which $\psi_4$ has the value (\ref{gr_psi4_closed}), with  
$a_1, a_2, b_1, b_2, c_1, c_2, d_1, d_2$ 
the eight complex parameters.  
\begin{table}
\caption{Free functions of the gravitational scalar potential}
\label{gr_consts}
\begin{center}
\begin{tabular}{| l | l |}
\hline
Function & Value \\
\hline
$a(r)$ & $\ a_1 r^2 (r-3M) + a_2$ \\
$b(r)$ & $\ b_1 r^2 +(r-M) b_2$ \\
$c(r)$ & \dis-\frac12 a_1(r-M)(r^2+4M^2)-\frac12 a_2+c_2(r-M)+c_1r^2$ \dis \phantom{ \frac x{\frac xx}}$ \\
$d(r)$ &\ \dis \frac12 b_1 r^2 +\frac12 b_2 r+d_1 r^2(r-3M)+d_2$\dis \phantom{ \frac x{\frac xx}}$\\
\hline
\end{tabular}
\end{center}
\end{table}

Now each $\Psi_H$ generates a vacuum perturbation for which the gauge invariant 
component $\psi_4$ vanishes.  It follows that $\psi_0$ vanishes as well, and a 
theorem of Wald \cite{Wald} restricts the perturbations associated with $\Psi_H$ to   
combinations of perturbations that change the mass and spin of the black hole, 
perturb Schwarzschild to a C metric or to Kerr-NUT, or are pure gauge.  We begin the 
next section by characterizing in this way the perturbation associated with each 
of the eight parameters of $\Psi_H$.

\subsection{Completion of the solution}

\noindent
{\em Fields corresponding to $\Psi_H$ } 

We will need only real values of the parameters $a_1, \ldots , d_2$, but will 
also include complex values of $a_2$, because Im($a_2$) corresponds to a 
perturbation taking Schwarzschild to Kerr, adding 
angular momentum to the Schwarzschild geometry.
  
We note first that the perturbed metric associated with $a_1$ is a C-metric perturbation.
The components of the perturbed metric and Weyl tensor in the radiation gauge have the following forms:    
\bsube
\begin{eqnarray}
h_{22}&=&-a_1A\, 2(r-3M)\cos\theta,\\
h_{23}&=&a_1A\, \frac{4M^2(r-2M)-r^3\sin^2\theta}{\sqrt2\ r^2\sin\theta},\\
h_{33}&=&a_1A\, \frac{2 M\left(r\sin^2\theta-4M\right)\cos\theta}{r\sin^2\theta}, 
\label{ha1}\eea\esube
\bsube\bea
\psi_0&=&\psi_2=\psi_4=0,\\
\psi_1&=&a_1A\, \frac{6\sqrt{2}M^3}{r^4\sin\theta },\\
\psi_3&=&-a_1A\, \frac{3M\sin\theta}{2\sqrt2\ r^2}.
\end{eqnarray}\label{psia1}\esube
To verify that this is a perturbed C metric, we refer in Wald \cite{Wald} to the perturbed 
tetrad produced by what he calls $\dot{p}$.  Using this perturbed tetrad to compute the 
perturbation in the metric $g_{\alpha\beta}=2n_{(\mu}l_{\beta)}-2m_{(\mu}{\overline m}_{\beta)}$, we calculate a perturbed metric.  It is then straightforward to find a gauge transformation from 
Wald's gauge to our radiation gauge.

When $a_2$ is real, the perturbation associated with it is pure gauge: 
The nonzero components of the perturbed metric,
\bsube\label{ha2}\bea
h_{22}&=&-a_2A\,\frac{2\cos\theta}{r^2},\\
h_{23}&=&a_2A\,\frac{\sqrt{2}\,\sin\theta}{r^2},\\
h_{33}&=&0,
\eea
\esube
and Weyl tensor,
\bsube\label{psia2}\bea
\psi_1&=&-a_2A\,\frac{ 3\sin\theta}{\sqrt2\ r^4},\\
\psi_2&=&a_2A\,\frac{3\cos\theta }{r^4},\\
\psi_3&=&a_2A\,\frac{3\sin\theta }{2\sqrt2\ r^4},
\eea\esube
are the components of $\Lie_\xi g_{\alpha\beta}$ and $\Lie_\xi C_{\alpha\beta\gamma\delta}$ for the gauge vector
\bsube\label{xia2}\bea
\xi_1&=&a_2A\,\frac{\cos\theta}{M},\\
\xi_2&=&-a_2A\,\frac{(r+2M)\cos\theta}{2M r},\\
\xi_3&=&-a_2A\,\frac{\sin\theta}{\sqrt{2}M}.
\eea\esube

For imaginary $a_2$, the perturbation is no longer pure gauge. Instead 
it describes a change in the angular momentum of the black hole, given by   
\be 
\dot J = ia_2 A.
\ee
The perturbed metric has components
\bea
h_{23} &=& - i\sqrt{2}\dot J \frac{\sin\theta}{r^2}, \\
h_{22} &=& h_{33} = 0.
\label{hkerr}\eea 

We show that this is a perturbation to Kerr, written in a radiation gauge,  
by exhibiting a gauge transformation to a gauge associated 
with Boyer-Lindquist coordinates.  Denote by $g^{\rm kerr}_{\alpha\beta}(M,J)$ the 
Kerr metric of mass $M$ and angular momentum $J$.  In 
Boyer-Lindquist coordinates, a perturbation 
\dis h^{\rm kerr}_{\alpha\beta} = \frac d{d\zeta} g^{\rm kerr}_{\alpha\beta}[M,J(\zeta)]_{\zeta=0}$ 
(with $J(0)=0$) has as its only nonzero component 
\be
 h^{\rm kerr}_{t\phi} = -2\frac{\dot J}r \sin\theta;
\ee
the corresponding nonzero components along the Kinnersley tetrad are 
\be 
h^{\rm kerr}_{13} = 2 h^{\rm kerr}_{23}= -i\sqrt2\,\dot J \, \frac{\sin\theta}\Delta .
\ee 
With the gauge vector $\xi^\alpha$ whose nonzero components are\\
\be
 \xi^3 =-\xi^4= \frac i{\sqrt2}\frac{\dot J}M \left[1+\frac r{2M}\ln\left(1-\frac{2M}r\right)\right], 
\ee
the perturbed metric $h_{\alpha\beta}= h^{\rm kerr}_{\alpha\beta}+\Lie_\xi g_{\alpha\beta}$ is the  
metric of Eq.~(\ref{hkerr}), as claimed.

The perturbation associated with $b_1$ is a change in the mass of the black hole of 
magnitude $\dot M =  3MAb_1$.  In the radiation gauge, metric and Weyl tensor perturbations have components    
\bsube\label{hb1}\bea
h_{22}&=&-2 b_1 A,\\
h_{23}&=&0,\\
h_{33}&=&b_1 A\frac{2\left(1+\cos^2\theta\right) }{\sin^2\theta},
\eea\esube
\vspace{-8mm}
\bsube\label{psib1}\bea
\psi_0&=&\psi_1=\psi_2=\psi_3=\psi_4=0.
\eea\esube
To verify that this is a perturbation of the Schwarzschild mass, we exhibit 
a gauge transformation to a Schwarzschild gauge, a gauge in which 
\dis h_{tt} = \frac d{d\zeta}\left(1-\frac{2M(\zeta)}r\right) = -\frac{2\dot M}r$, 
\dis h_{rr}= -\frac d{d\zeta}\left[\left(1-\frac{2M(\zeta)}r\right)^{-1}\right]
= -\frac{2\dot M}r\frac{r^4}{\Delta^2}$, with all other coordinate components vanishing.
The corresponding nonzero tetrad components of $h_{\alpha\beta}$ in this gauge are  
\be
h_{11} = -\frac{4\dot M}r\frac{r^4}{\Delta^2}, \qquad h_{22} = - \frac{\dot M}r.
\label{mdot}\ee
The gauge transformation that yields a perturbed metric of this form is generated 
by the vector $\xi^\alpha$ with components
\bsube\label{xib1}\bea
\xi_1&=& b_1 A\, \frac{r}{\Delta}\left[r^2-4Mr -8M^2 - t(r-2M)+8M(r-2M)\ln(r-2M)\right],\\
\xi_2&=&b_1 A\, \frac{1}{2r}\left[3r^2-4Mr -8M^2 - t(r-2M)+8M(r-2M)\ln(r-2M)\right],\\
\xi_3&=&-b_1 A\, \sqrt{2}r \cot\theta.
\eea\esube
The corresponding gauge perturbation is given by 
\bsube\label{gaugeb1}\bea
(\Lie_\xi g)_{11}&=&b_1A\frac{12  M r^3}{\Delta^2},\\
(\Lie_\xi g)_{22}&=&-b_1 A\frac{2r-3M}{r},\\
(\Lie_\xi g)_{33}&=&b_1A\frac{2\left(1+\cos^2\theta\right) }{\sin^2\theta}.
\eea\esube
Finally, the gauge-transformed metric obtained by subtracting (\ref{gaugeb1}) 
from (\ref{hb1}) has the form (\ref{mdot}), with $\dot M =  3MAb_1$, 
as claimed.

 The perturbation associated with $b_2$ is again a change in 
the mass of the black hole, in this case with $\dot M=   Ab_2$. In the 
radiation gauge, it has the form 
\bsube\label{hb2}\bea
h_{22}&=&- b_2 A\, \frac{2(r-M)}{r^2},\\
h_{23}&=&-b_2 A\, \frac{\sqrt{2} (r-2M) \cos\theta}{r^2\sin\theta },\\
h_{33}&=&b_2 A\, \frac{2\left(1+\cos^2\theta\right) }{r \sin^2\theta},
\eea\esube
\bsube\label{psib2}\bea
\psi_0&=&\psi_3=\psi_4=0,\\
\psi_1&=&-b_2 A\, \frac{3 \sqrt{2} M \cos\theta}{r^4\sin\theta },\\
\psi_2&=&b_2 A\, \frac{r-3M}{r^4}.
\end{eqnarray}\esube
A gauge transformation to the Schwarzschild gauge (\ref{mdot}) is in this 
case generated by the gauge vector
\bsube\label{xib2}\bea
\xi_1&=&-b_2 A\, \frac{r}{\Delta}\left[r+2M-2(r-2M)\ln\left(\frac{r-2M}{\sin\theta}\right)\right],\\
\xi_2&=&b_2 A\, \frac{ \Delta}{2r^2}\left[1+2\ln\left(\frac{r-2M}{\sin\theta}\right)\right],\\
\xi_3&=&-b_2 A\, \sqrt{2}\cot\theta, 
\end{eqnarray}\esube
for which
\bsube\label{gaugeb2}\bea
(\Lie_\xi g)_{11}&=&b_2 A\, \frac{4r^3}{\Delta^2},\\
(\Lie_\xi g)_{22}&=&-b_2 A\, \frac{\Delta}{r^3},\\
(\Lie_\xi g)_{23}&=&-b_2 A\, \frac{\sqrt{2} (r-2M) \cos\theta}{r^2\sin\theta},\\
(\Lie_\xi g)_{33}&=&b_2 A\, \frac{2\left(1+\cos^2\theta\right)}{r \sin^2\theta}.
\end{eqnarray}\esube
Subtracting (\ref{gaugeb2}) from (\ref{hb2}) yields the form (\ref{mdot}), with $\dot M =  Ab_2$.

The remaining perturbations are all pure gauge.  For each 
parameter, we list below the components of the associated metrics, the components of the perturbed Weyl tensor, and the nonzero components of the gauge vector. 

\noindent
$c_1$:
\bsube\label{hc1}\begin{eqnarray}
h_{22}&=&h_{23}=0,\\
h_{33}&=&c_1 A\,\frac{4 \cos\theta}{\sin^2\theta},
\eea\esube
\vspace{-1cm}
\bsube\label{psic1}\begin{eqnarray}
\psi_0&=&\psi_1=\psi_2=\psi_3=\psi_4=0,
\end{eqnarray}\esube
\be
\xi_3 = -c_1 A \,\frac{\sqrt{2}r}{\sin\theta}.
\label{xic1}\ee

\noindent
$c_2$:
\bsube\bea
h_{22}&=&0,\\
h_{23}&=&-c_2 A\,\frac{\sqrt{2}\, (r-2M)}{r^2\sin\theta},\\
h_{33}&=&c_2 A\,\frac{4 \cos\theta}{r\sin^2\theta},
\eea\esube
\vspace{-1cm}
\bsube\bea
\psi_0&=&\psi_2=\psi_3=\psi_4=0,\\
\psi_1&=&-c_2 A\,\frac{3 \sqrt{2}M}{r^4\sin\theta},
\end{eqnarray}\esube
\bsube\bea
\xi_1&=&c_2 A\,2  \ln \left(\cot\frac{\theta}{2}\right),\\
\xi_2&=&c_2 A\,\frac{\Delta}{r^2}\ln \left(\cot\frac{\theta}{2}\right),\\
\xi_3&=&-c_2 A\,\frac{\sqrt{2}}{\sin\theta}.
\end{eqnarray}\esube

\noindent
$d_1$:
\bsube\bea
h_{22}&=&h_{23}=0,\\
h_{33}&=&-d_1A\,\frac{12 M}{\sin^2\theta},
\eea\esube
\vspace{-1cm}
\bsube\bea
\psi_0&=&\psi_1=\psi_2=\psi_3=\psi_4=0,
\end{eqnarray}\esube
\bea
\xi_3&=& d_1A\,3\sqrt{2} M r \left( \cot\theta + i \phi \sin\theta \right).
\end{eqnarray}

\noindent
$d_2$:
\begin{eqnarray}
h_{22}&=&h_{23}=h_{33}=0,\\
\psi_0&=&\psi_1=\psi_2=\psi_3=\psi_4=0.
\end{eqnarray}

\noindent
{\em Fields corresponding to $\Psi_P$ }

The perturbed metric associated with $\Psi_P$ has the nonzero components 
\bsube
\begin{eqnarray}
h_{22}&=&-(\bar{\delta}+2\bar\beta)(\bar{\delta}+4\bar\beta)\bar{\Psi}_P
	 -(\delta + 2\beta)(\delta + 4\beta) \Psi_P\nonumber\\
&=&  A \frac{2}{r^2R}
   \left[(\mfr^2+\mfr_0^2)M^2+\mfr^2\mfr_0^2-M^4-4M^2\mfr\mfr_0\cos\theta
   +2M^4\cos^2\theta \right],\\
h_{23} &=& (D+\rho) (\delta + 4\beta){\Psi}_P +(D+\rho) (\bar{\delta} + 4\bar\beta)\bar{\Psi}_P\nonumber\\
&=& A \frac {\sqrt{2}}{r^2R\sin\theta }
    \left\{(2\mfr-M)r \mfr_0 M-(\mfr-M)\mfr_0^3+
    (\mfr-2M)[r M^2+(\mfr-M)\mfr_0^2]\cos\theta\right.\nonumber\\
   &&\left.-2(2\mfr-M)\mfr_0 M^2\cos^2\theta+2M^4\cos^3\theta\right\},\\
h_{3 3} & =& (D - \rho)(D + 3\rho)\bar\Psi_P + (D - \rho)(D + 3\rho)\Psi_P\nonumber\\
& =&-A\frac{2}{rR\sin^2\theta}
   \left\{\mfr \mfr_0^2-M^2\mfr-M(2\mfr^2+\mfr_0^2)+M^3+2\mfr_0(M^2+2M
   \mfr-\mfr^2-\mfr_0^2)\cos\theta\right.\nonumber\\
   &&\left.[(3\mfr-M)\mfr_0^2+(\mfr-M)M^2]\cos^2\theta-2\mfr_0M^2\cos^3\theta\right\}.
\end{eqnarray}
\esube

We now compute the change in the black-hole area for the perturbed metric 
associated with $\Psi_P$.  
Because the perturbed solution is static, the horizon is again a Killing horizon.  Because the identification of perturbed and unperturbed spacetimes has been chosen to keep $t^\alpha$ as the Killing vector of the perturbed spacetime, the horizon is outermost set of points where $t^\alpha$ 
is null, where 
$
 t^\alpha t_\alpha = g_{tt}= 0.
$ 
Then, using $h_{t t}=h_{22}$ (this is a consequence of the gauge, $h_{\alpha\beta}l^\beta=0$, and the definition of the tetrad) , we obtain the perturbation $\dot r$ in the 
horizon radius at fixed $\theta$ and $\phi$ by writing    
\begin{eqnarray}\label{gr_horizon_eqn}
0&=& \frac d{d\zeta}g_{tt}[\zeta,r(\zeta)]
  = \frac d{d\zeta}\left[1-\frac{2M}{r(\zeta)}\right] + \left. h_{22}\right|_{r=2M} 
  =\left.\frac{2M}{r^2}\right|_{r=2M}\dot r +\left.h_{22}\right|_{r=2M} 
\nonumber\\
&=&  \frac 1{2M}\dot r + A[r_0-M(1+\cos\theta)], 
\eea
whence
\be
\dot r = -2AM[r_0-M(1+\cos\theta)].
\ee

The change in the area of the horizon has, in general, contributions 
from the change $\dot r$ in its position and the change in the 
element of area, 
\dis \frac d{d\zeta}\sqrt{{}^2g}\,d\theta d\phi = \sqrt{{}^2g}\,(h_{\hat{\theta}\hat{\theta}}+h_{\hat{\phi}\hat{\phi}})d\theta d\phi$. 
In the radiation gauge, however, $h_{34}=\frac 1 2 (h_{\hat{\theta}\hat{\theta}}+h_{\hat{\phi}\hat{\phi}})=0$, and the change in 
the horizon area is given by  
\begin{eqnarray}
\dot A_{horizon}&=&\frac d{d\zeta} \int r_{horizon}^2 d\Omega = -2\int r_{horizon}\dot r d\Omega\\
&=&-32 \pi A M^2 (r_0-M).
\label{ahorizon}\end{eqnarray}

The local first law of black hole thermodynamcs relates the change in the 
Komar mass, \dis M_K=-(8\pi)^{-1}\int_S \nabla^\alpha t^\beta dS_{\alpha\beta}$, 
of a Killing horizon to the change in its area:\\
\dis\dot M=\frac1{32\pi M}\dot A_{horizon}$.  By Eq.~({\ref{ahorizon}), the 
particular solution associated 
with $\Psi_P$ changes the mass of the horizon by  
\be 
 \dot M = -AM(r_0-M).
\label{dotmk}\ee
We can thus keep area and mass of the horizon constant by adding 
to $h_{\alpha\beta}[\Psi_P]$ the 
$l=0$ metric perturbation (\ref{mdot}) associated with a change 
\be 
 \dot M = AM(r_0-M)
\label{dotm}\ee
of the Schwarzschild mass.

Note that instead of computing the change in area, we could have directly computed the change in the horizon's Komar mass.  

As mentioned earlier, in addition to the singular radiation-gauge forms 
of the mass perturbation associated with $b_1$ and $b_2$, there is a 
radiation gauge in which the mass perturbation is smooth on the 
exterior Schwarzschild spacetime. Choosing a gauge vector $\xi^\alpha$ 
with components \dis\xi_2 = \frac{2M}r - \frac\Delta{r^2}\ln\left(\frac r{2M}-1\right),\ \xi_1 = 2\frac{r^2}\Delta \xi_2$,  we obtain from Eq. (\ref{dotm}) a perturbed metric $h_{\alpha\beta}\equiv h^S_{\alpha\beta}-\Lie_\xi g_{\alpha\beta}$
whose only nonzero component is \dis h_{22} = -\frac 2r$.

\subsection{Singularity}

For the static charge of the last section and for a static mass on a flat background (treated 
in the next section) the perturbed electromagnetic and gravitational fields are singular 
only at the position of the particle, although any choice of radiation gauge yields components of 
$A_\alpha$ and $h_{\alpha\beta}$, respectively, that 
are singular on (at least) a ray whose endpoint is the particle.  For a static mass in a 
Schwarzschild background, we have seen that the perturbed gauge-invariant components of the 
Weyl tensor are also nonsingular except at the particle.  This is surprising, because the 
metric perturbation must have an additional singularity that can be regarded as a string or 
strut supporting the particle. An initially static mass with no 
external support will fall inward: If the linearized Einstein equations are satisfied 
for a source of the form $\dot T^{\alpha\beta} = \dot\rho u^\alpha u^\beta$, with 
$u^\alpha$ along the timelike Killing vector that Lie derives $\delta\rho$, then the perturbed 
Bianchi identities imply 
\be 
   0 = \nabla_\beta \dot T^{\alpha\beta} = \dot\rho u^\beta\nabla_\beta u^\alpha.
\ee
It follows that the particle moves along a geodesic of the background geometry, 
$u^\beta\nabla_\beta u^\alpha=0$, contradicting the assumption that the particle is static.   
We will find that the perturbed geometry 
has a conical singularity, and it can be chosen to run along a radial line from the particle 
to infinity.  There is also, associated with any radiation gauge, a singularity of the perturbed 
metric that, like that of the vector 
potential in the last section, lies along a ray with endpoint at the particle and 
that can be removed by a gauge transformation.

We now show that one can choose the parameters $a_1, \ldots d_2$ to make the components of 
the metric finite either on a radial ray above the particle ($r>r_0$) or below the particle
($r<r_0$), but not both.  If one departs from a radiation gauge, one can make the metric finite above and below the particle, but one cannot avoid a conical singularity.

Although the most efficient way to obtain necessary and sufficient conditions for smoothness of the perturbed metric is again to use the Hertz potential, in the 
case at hand the perturbed metric will not be smooth.   We can at best 
find a gauge in which its components are finite, and, for clarity of presentation, 
we will use the direct expansion of the components of $h_{\alpha\beta}$.  

The components of the perturbed metric near $\theta=0$, are given for $r<r_0$ by
\bsube\begin{eqnarray}
h_{33}&=&A\frac{4}{r \theta^2}
 \left[-2M^2 a_1 +b_2+c_2+ r_0^2+(b_1+c_1-3Md_1-r_0)r\right]\nonumber\\
& & -A\frac{2}{3 r}
  \left\{-2M^2 a_1+b_2+c_2+r_0^2\right.\nonumber\\
&&\left. \phantom{xxxxxx}+[3M(-a_1+3d_1+1)+b_1+c_1-3Md_1-r_0]r\right\}+O(\theta^2),\\
h_{23}&=&A\frac{\sqrt{2}(r-2M)}{r^2\theta}\left(2M^2a_1-b_2-c_2-r_0^2\right)+O(\theta^0);
\end{eqnarray}\esube
and for $r>r_0$ by 
\bsube\begin{eqnarray}
h_{33}&=&A\frac{4}{r \theta^2}
 \left[-2M^2 a_1 +b_2+c_2-r_0^2+(b_1+c_1-3Md_1+r_0)r\right]\nonumber\\
& & -A\frac{2}{3 r}
  \left\{-2M^2 a_1+b_2+c_2-r_0^2\right.\nonumber\\
&&\left. \phantom{xxxxxx}+[3M(-a_1+3d_1-1)+b_1+c_1-3Md_1+r_0]r\right\}+O(\theta^2),\\
h_{23}&=&A\frac{\sqrt{2}(r-2M)}{r^2\theta}\left(2M^2a_1-b_2-c_2+r_0^2\right)+O(\theta^0).
\eea\esube

Then $h_{\alpha\beta}$ is smooth at $\theta=0$ if and only if the coefficients of 
$\theta^{-2}$ and $\theta^0$ in $h_{33}$ and the coefficient of $\theta^{-1}$ in $h_{23}$ vanish. 
The corresponding conditions on the parameters are 
\bsube\bea
 2 M^2 a_1 -b_2-c_2&=& r_0^2, \quad b_1+c_1-3M d_1 =  r_0, \quad a_1-3d_1 =  1, \quad r<r_0; \label{hsmooth0<} \\
 2 M^2 a_1 -b_2-c_2&=& - r_0^2, \quad b_1+c_1-3M d_1 =  - r_0, \quad a_1-3d_1 = -  1, \quad r>r_0 .
\label{hsmooth0>}\eea\esube
Because the conditions for $r<r_0$ are not consistent with those for $r>r_0$, one 
cannot find an single radiation gauge smooth at $\theta=0$ both above and below the particle.

Near $\theta=\pi$, the perturbed metric has components
\bsube\bea
h_{33}&=&A\frac{4}{r (\pi-\theta)^2}
      \left[2M^2a_1+b_2-c_2-(r_0-2M)^2+(b_1-c_1-3M d_1-r_0 +2M)r\right]\nonumber\\
& &-A\frac{2}{3 r}\{2M^2a_1+b_2-c_2-(r_0-2M)^2+[b_1-c_1-3Md_1-r_0+2M +3M(a_1+ 3d_1-1)]r\}\nonumber\\ & &+O\left[(\pi-\theta)^2\right],\\
h_{23}&=& A \frac{\sqrt{2}(r-2M)}{r^2(\pi-\theta)}\left[2M^2a_1+b_2-c_2-(r_0-2M)^2\right]+O[(\pi-\theta)].
\end{eqnarray}\esube

Then $h_{\alpha\beta}$ is smooth at $\theta=\pi$ if and only if the coefficients of 
$(\pi-\theta)^{-2}$ and $(\pi-\theta)^0$ in $h_{33}$ and the coefficient of $(\pi-\theta)^{-1}$ in $h_{23}$ vanish. 
The corresponding conditions on the parameters are 
\bea
 2 M^2 a_1 +b_2-c_2&=& (r_0-2M)^2, \quad b_1-c_1-3M d_1 =  (r_0-2M), \nonumber\\
a_1+ 3d_1= 1.
\label{hsmoothpi}\eea

Requiring the perturbed metric to be either\\
(I) smooth everywhere except $\theta=0, \ r>r_0$, or\\
(II) smooth everywhere except $\theta=0, \ r<r_0$\\
uniquely determines the subset of parameters 
$\{a_1, b_1, b_2, c_1, c_2, d_1\}$. For the two cases, these parameters -- 
the unique solutions to Eqs.~(\ref{hsmoothpi}) and to either Eqs.~(\ref{hsmooth0<}) or  
(\ref{hsmooth0>}) -- are listed in Table \ref{h_smooth}.
\begin{table}[h]
\caption{Singularity location for two choices of parameters}
\label{h_smooth}
\begin{center}
\begin{tabular}{| c | c | c |}
\hline
Parameter  & (I) Singularity at &(II) Singularity at\\
 & $\theta=0$, $r\geq r_0$ & $\theta=0$, $r\leq r_0$\\
\hline
$a_1$ & $ 1$ & $0$\\
$a_2$ & 0& 0\\
$b_1$ & $ (r_0-M)$ & $0$\\
$b_2$  & $-2M(r_0-M)$ & $ r_0^2-2Mr_0+2M^2$\\
$c_1$  &$ M$ & $-r_0+M$ \\
$c_2$  & $-r_0(r_0-2M)$ & $ 2M(r_0-M)$ \\
$d_1$ & $0$ & $ \frac{1}{3}$\\
\hline
\end{tabular}
\end{center}
\end{table}
The parameter $a_2$ does 
not affect the singular structure of $h_{\alpha\beta}$ and is arbitrary; for 
definiteness it is set to zero in the table. The trivial parameter 
$d_2$ does not alter the metric and is not listed. 

The perturbed metric $h^{(I)}_{\alpha\beta}$ corresponding to the 
parameters of the first column 
is, in particular, smooth for $r<r_0$.  The nonzero values of the parameters 
$b_1$ and $b_2$ correspond to a perturbation of the Schwarzschild mass given by 
\be
\dot M =  3AMb_1 +Ab_2 = AM(r_0-M).
\label{dotm1}\ee
This is exactly the change in mass (\ref{dotm}) needed to cancel the change 
in black hole area and mass of the particular solution.  Smoothness 
of a radiation gauge in the region $2M<r<r_0$ between the horizon and the particle 
thus requires that the perturbation involve no change in the black hole mass.
Just as the vector potential in a radiation gauge is smooth on a sphere only if 
the sphere encloses no perturbed charge, our result suggests that the perturbed 
metric can be smooth on a sphere only if the sphere encloses no perturbed mass.

We thus expect that for the perturbed metric $h^{(II)}_{\alpha\beta}$, 
smooth for $r>r_0$ (and asymptotically flat), the asymptotic mass 
must vanish, and direct calculation 
verifies this.  For a flat background, we will see in the next section 
that one can obtain from $h^{(II)}_{\alpha\beta}$ a perturbed metric that agrees up to 
gauge with $h^{(II)}_{\alpha\beta}$ by adding an $l=0$ part, a perturbed 
Schwarzschild solution centered at the origin that restores the total asymptotic 
mass.  For a Schwarzschild background, however, there is a second physical 
difference between 
solutions (I) and (II): The difference in the values of $a_1$ means that 
$h^{(I)}_{\alpha\beta}$ and $h^{(II)}_{\alpha\beta}$ differ by a perturbed C metric.  
The solution that agrees up to gauge with $h_{\alpha\beta}^{(I)}$ has the 
form $h^{(II)}_{\alpha\beta}+ h_{\alpha\beta}^{\rm Schwarzschild}
+h_{\alpha\beta}^{C metric}$.  The C-metric term has a conical singularity on 
the axis; adding it in the gauge of \cite{Kinnersley}, and adding the Schwarzschild 
perturbation in a Schwarzschild gauge yields a perturbed metric   
that is finite but not smooth on the axis.

\section{Flat Space}
\label{mass2}

When the mass $M$ of the background Schwarzschild geometry is set to zero, the perturbation 
is just the gravitational field of a point mass linearized about a flat background: the linearized 
Schwarzschild solution.  Like the electromagnetic field of a point charge in a Schwarzschild 
background, the perturbed geometry is now singular only at the position of the particle, 
and any other singularity in the tensor $h_{\alpha\beta}$ is an artifact of the choice of gauge.

The origin of coordinates inherited from the $M\neq 0$ case is displaced from the particle by a distance $r_0$; and the inherited tetrad is aligned with inward and outward radial null geodesics associated with the spatial origin, not with the particle.  With respect to a tetrad associated with the null geodesics that start or end {\em at the particle}, the perturbation is algebraically special, having as the only nonzero component of its perturbed Weyl tensor \dis\psi_2= -\frac{\mathfrak m}{R^3}$, with $R$ the distance to the particle.  With respect 
to our null tetrad, however, the perturbation is not algebraically special: Its components are 
\bsube\bea
\psi_0&=& -3\frac{\mathfrak m}{R^5}r_0^2\sin^2\theta,	\\
\psi_1 &=&  \frac3{\sqrt2}\frac{\mathfrak m}{R^5}r_0(r-r_0\cos\theta)\sin\theta,	\\ 
\psi_2 &=& - \frac{\mathfrak m}{R^5}
	\left(R^2-\frac32 r_0^2\sin^2\theta\right),	\\
\psi_3 &=& - \frac3{2\sqrt 2}\frac{\mathfrak m}{R^5}r_0(r-r_0\cos\theta)\sin\theta,	\\
\psi_4 &=& -\frac34\frac{\mathfrak m}{R^5}r_0^2\sin^2\theta,	
\eea\esube
where 
\be R= \sqrt{r^2+r_0^2-2rr_0\cos\theta}.
\ee

(This coordinate expression for the distance $R$ to 
the particle is the $M=0$ limit of the length $R$ of the last section.)  
In contrast to perturbations of Kerr and Schwarzschild, where only $\psi_0$ and $\psi_4$ are 
gauge invariant, the full Weyl tensor is gauge invariant, because the Weyl tensor  
of the flat background vanishes.  The corresponding components along the null tetrad 
$l^\alpha,n^\alpha,m'^\alpha$ differ from these only by a 
phase and are manifestly smooth except at $R= 0$, the position of the particle, and at $r= 0$, 
where the tetrad is singular.  
\bsube\bea
\psi_0'&=& - 3\frac{\mathfrak m}{R^5}r_0^2\sin^2\theta e^{2i\phi} 
= - 3\frac{\mathfrak m}{R^5}\frac{r_0^2}{r^2}(x+iy)^2,	\\
\psi_1' &=&  \frac3{\sqrt2}\frac{\mathfrak m}{R^5}r_0(r-r_0\frac zr)\sin\theta e^{i\phi}
		= \frac3{\sqrt2}\frac{\mathfrak m}{R^5}\frac{r_0}r(r-\frac{r_0}r z)(x+iy), \\ 
\psi_2' &=& - \frac{\mathfrak m}{R^5}
	\left(R^2-\frac32 r_0^2\sin^2\theta\right)
	=- \frac{\mathfrak m}{R^5}
	\left[R^2-\frac32 \frac{r_0^2}{r^2}(x^2+y^2)\right],	\\
\psi_3' &=& -\frac3{2\sqrt 2}\frac{\mathfrak m}{R^5}r_0(r-r_0\cos\theta)\sin\theta e^{-i\phi}
		= - \frac3{2\sqrt 2}\frac{\mathfrak m}{R^5}\frac{r_0}r(r-\frac{r_0}r z)(x-iy),\\
\psi_4' &=& - \frac34\frac{\mathfrak m}{R^5}r_0^2\sin^2\theta
	= -\frac34\frac{\mathfrak m}{R^5}\frac{r_0^2}{r^2}(x-iy)^2.	
\eea\esube
The components $\psi_i'$ are smooth at $\theta = \pi$, although the tetrad is not; this is due  
to the fact that the tetrad $l^\alpha, n^\alpha,e^{-i\phi}m^\alpha $ is smooth at $\theta = \pi$, 
and the components of the Weyl tensor associated with that tetrad are simply $\bar\psi_i'$.

The perturbed mass is given by 
\begin{eqnarray}
\lim_{r\to\infty}\left(-r^3\psi_2\right)
&=&\lim_{r\to\infty}r^3\frac{\mathfrak m}{R^5}
	\left(R^2-\frac32 r_0^2\sin^2\theta\right)\nonumber\\
&=&{\mathfrak m}.
\end{eqnarray} 

We will again find the perturbed metric $h_{\alpha\beta}$ in a radiation gauge, choosing, within the 
family of radiation gauges, one for which the metric components are finite except on the $\theta=0$ 
axis for $r>r_0$ or for $r<r_0$. As was the case for the vector potential of the smooth 
electromagnetic field of a point particle, one cannot globally choose parameters $a_1, \ldots d_2$ for which the metric is smooth and regular everywhere outside the particle.  

As in the previous sections, if the Hertz potential  $\Psi'$ associated with
the smooth tetrad $l^\alpha,n^\alpha,m'^\alpha$ is smooth 
at $\theta = 0$, then the perturbed metric associated with this tetrad 
is smooth at $\theta=0$.     

From Eqs.~(\ref{PsiP}) and (\ref{PsiH}), a particular solution $\Psi_P$ associated 
with the Kinnersley tetrad has the form 
\be 
\Psi_P = -\frac23 A \frac{R^3}{\sin^2\theta}, 
\label{psipflat}\ee
and the general homogeneous solution has the form
\bea 
\Psi_H &=&  A\frac1{3\sin^2\theta} \left[(a_1\cos^3\theta-3a_1\cos\theta+6d_1)r^3 
			+ 3(b_1\cos^2\theta+b_1+2c_1\cos\theta)r^2\right.\nonumber\\
&&\left.		\hspace{5mm}+(a_2\cos^3\theta-3a_2\cos\theta+6d_2)
		      \  + \ 3(b_2\cos^2\theta+b_2+2c_2\cos\theta)r 
		       \right].
\label{psihflat}\eea 
The Hertz potentials associated with tetrads smooth at $\theta = 0$ and 
$\theta=\pi$ are then $\Psi e^{\pm 2i\phi}= (\Psi_P+\Psi_H)e^{\pm 2i\phi}$.

We will show that one can make the corresponding metric $h_{\alpha\beta}$ smooth at 
$\theta=\pi$, by a choice of parameters that correspond to gauge transformations. 
One can simultaneously choose parameters to make $h_{\alpha\beta}$ smooth at 
$\theta = 0$ for either $r\geq 0$ or for $r\leq 0$. But no consistent choice of 
parameters allows the perturbed metric to be smooth both above and below the 
particle on the ray through the particle. Moreover, smoothness at $\theta=0$ 
above the particle requires a choice of parameters that alters the asymptotic 
mass; to recover for $r>r_0$ the metric of a static particle, one must add
to the radiation-gauge solution an $l=0$ part, a linearized Schwarzschild solution 
centered at the origin.          

We begin by describing changes in the perturbed metrics 
associated with the parameters $a_1, \ldots, d_2$ that arise for a flat background.
For Kerr geometries with nonzero mass, Wald \cite{Wald} shows 
that any perturbation with $\psi_0$ or $\psi_4=0$ is a type D perturbation or 
is pure gauge.  Because the proof relies on the fact that a gauge transformation can make  
$\psi_1$ and $\psi_3$ vanish, it fails for a flat background.  There are, in fact, 
vacuum perturbations of flat space for which the gauge invariant components 
$\psi_1, \psi_2$ and $\psi_3$ have nonzero values, although $\psi_0=\psi_4=0$. In 
particular, we will see that the perturbation associated with $a_2$ is no longer 
pure gauge but is a perturbation of this kind. (This can happen because the gauge vector
(\ref{xia2}) has no $M=0$ limit). 

Conversely, we will see that the perturbations associated with $a_1$ and $b_2$ are 
now pure gauge, although they are C metric and mass perturbations when $M\neq 0$.         
The components of the perturbed metric and Weyl tensor associated with each 
of the parameters can be read off from the 
corresponding expressions in the last section by setting $M$ to zero. 
Because each parameter corresponds to a vacuum perturbation and the background 
Riemann tensor vanishes, the perturbation is pure gauge if and only if the 
components of the perturbed Weyl tensor vanish. That is, a perturbation of 
flat space is pure gauge iff the perturbed Riemann tensor 
$\dot R^\alpha_{\ \beta\gamma\delta}$ vanishes; and, for a perturbation of flat space, we have \\
\centerline{\dis
\dot R^\alpha_{\ \beta\gamma\delta}= 0\Leftrightarrow 
\dot R_{\alpha\beta\gamma\delta}= 0 \Leftrightarrow
\dot R_{\alpha\beta}=0 \mbox{ and } \dot C_{\alpha\beta\gamma\delta}=0$.}
 
Eqs.~(\ref{psia1}) imply that the perturbed Weyl tensor associated with $a_1$ vanishes, 
and thus that $a_1$ corresponds to a gauge perturbation,  
$h_{\alpha\beta}=\Lie_\xi g_{\alpha\beta}$.

Writing \dis\widetilde A \equiv A|_{M=0}= \frac{\mathfrak m}{r_0^2}$, we obtain 
\bsube
\begin{eqnarray}
h_{22}&=&-a_1\widetilde A\  2r\cos\theta,\\
h_{23}&=&-a_1\widetilde A\  \frac1{\sqrt2}r\sin\theta,\\
h_{33}&=&0, 
\label{ha1flat}\eea\esube
\bsube\bea
\xi_1&=&a_1\tilde A\, \frac12 (t-r)^2\cos\theta,\\
\xi_2&=&-a_1\tilde A\, \frac14(t-r)(t+3r)\cos\theta,\\
\xi_3&=&-a_1\tilde A\, \frac1{2\sqrt2}(t^2-r^2)\sin\theta.
\end{eqnarray}\label{xia1flat}\esube

Because the components (\ref{psia2}) of the perturbed Weyl tensor associated with $a_2$ are nonzero, the perturbation associated with $a_2$ is no longer pure gauge. 

The perturbation associated with $b_1$ was a change of mass proportional to $M$; 
because Eqs. (\ref{psib1}) imply $\psi_i=0$, it is pure gauge when $M=0$:    
\bsube\begin{eqnarray}
h_{22}&=&-2 b_1\tilde A, \\
h_{23}&=&0,\\
h_{33}&=&b_1 \tilde A\ \frac{2\left(1+\cos^2\theta\right) }{\sin^2\theta},
\end{eqnarray}\esube
\bsube
\begin{eqnarray}
\xi_1&=&b_1 \tilde A\ (r-t),\\
\xi_2&=&b_1 \tilde A\ \frac{1}{2} (3r-t),\\
\xi_3&=&\!-b_1 \tilde A\ \sqrt{2}\,  r \cot\theta.
\end{eqnarray}\esube

The metric associated with $b_2$ remains a mass perturbation, a linearized 
Schwarzschild solution, centered at the origin with mass $-{\mathfrak m}b_2/r_0^2$.
Finally, the perturbations associated with $c_1$ and $c_2$ are again pure gauge; 
and the perturbations associated with {\em both} $d_1$ and $d_2$ now vanish
($h_{\alpha\beta}=0$). To summarize: In flat space the 
perturbations associated with $a_1,b_1,c_1,c_2$ are pure gauge, and those 
associated with $d_1$ and $d_2$ vanish.

Then one can find a smooth perturbed metric of a static particle in flat space from a
radiation-gauge Hertz potential only if one can choose the six parameters $a_1,b_1, c_1,c_2,d_1,d_2$
to make the perturbed metric smooth.    
Now $h_{\alpha\beta}$ is smooth on the axis if and only if its components in a smooth 
basis are smooth.  Because $h_{2'3'}$ and $h_{3'3'}$ have angular dependence 
$e^{im\phi}$ and $e^{2im\phi}$, they are smooth at $\theta = 0$ only if   
$h_{22}=O(\theta^0),h_{23'}=O(\theta), h_{3'3'}=O(\theta^2)$; and this is true if and only if 
\be
h_{22}=O(\theta^0), \quad h_{23}=O(\theta), \quad h_{33}=O(\theta^2).
\ee
 Similarly, the metric can be smooth 
at $\theta = \pi$ only if 
\be
h_{22}=O((\theta-\pi)^0),\quad h_{23}=O((\theta-\pi)), \quad h_{33}=O((\theta-\pi)^2).
\ee

These requirements impose more algebraic conditions on the parameters than 
can be simultaneously satisfied.  One can, however, choose parameters for 
which the metric is smooth everywhere except on a ray that does not intersect 
the particle (and at the particle itself): That is one can find a perturbed 
metric from a Hertz potential $\Psi'$ that is smooth everywhere except $\theta = \pi$.

As we saw with a Schwarzschild background, one can also choose parameters that make the 
perturbation regular in a radiation gauge either inside or outside the particle.  
Inside the particle, the choice is simply a gauge transformation.  Any choice of 
parameters that make the perturbation regular outside the particle require zero net 
mass for $r\geq r_0$. One can obtain a regular solution in a radiation gauge for 
the $l\neq 0$ part of the perturbation; to complete the solution one must add 
an $l=0$ part -- corresponding to a mass at the origin -- in a different gauge.  
 
One can find the allowed parameters either by directly expanding the metric for 
arbitrary parameter values near $\theta=0$ and $\theta = \pi$ or, more simply, by 
examining the Hertz potential near $\theta = 0$ and $\theta=\pi$.  Now $\Psi'$ is 
smooth near $\theta=0$ ($\theta = \pi$) if and only if $\Psi = O(\theta^2)$ 
($\Psi = O((\theta-\pi)^2)$).  From Eqs.~(\ref{psipflat},\ref{psihflat}), 
the parts $\Psi_P$ and $\Psi_H$ of the Hertz potential have near $\theta=0$ the forms 
\bsube\label{psiphat0}
\bea
\Psi_P &=&- \tilde A \frac23 |r-r_0|\left[(r-r_0)^2\theta^{-2}+\frac13(r-r_0)^2
	+\frac32 r_0r\right] + O(\theta^2),\\
\Psi_H &=& \tilde A\frac23\left\{\phantom{\frac12}\!\!\!\! 
    \left[(-a_1+3d_1) r^3 + 3(b_1+c_1)r^2 + 3(b_2+c_2)r - a_2+ 3d_2\right]\theta^{-2} \right.
\nonumber\\
&&+ \left.\left[\frac13(-a_1+3d_1) r^3-\frac12(b_1+c_1)r^2 - \frac12(b_2+c_2)r 
	+\frac13(-a_2+ 3d_2)\right]\right\} + O(\theta^2).\nonumber\\
\eea\esube
Then, for $\Psi= \Psi_P+\Psi_H$, we have 
\bsube\bea
\Psi &=& \tilde A\frac23\left\{\phantom{\frac12}\hspace{-4mm}
    \left[(-1-a_1+3d_1) r^3 + 3( r_0+b_1+c_1)r^2 + 3(-r_0^2+b_2+c_2)r +r_0^3- a_2+ 3d_2\right] \theta^{-2}\right.
\nonumber\\
&&+ \left.\left[\frac13(-1-a_1+3d_1) r^3-\frac12( r_0+b_1+c_1)r^2 - \frac12(-r_0^2+b_2+c_2)r 
	+\frac13(r_0^3-a_2+ 3d_2)\right] \right\}\nonumber\\
 &&+ O(\theta^2), \quad r>r_0,\\
\Psi &=& \tilde A\frac23\left\{\phantom{\frac12}\hspace{-4mm}
    \left[( 1-a_1+3d_1) r^3 + 3(-r_0+b_1+c_1)r^2 + 3( r_0^2+b_2+c_2)r -r_0^3- a_2+ 3d_2\right] \theta^{-2}\right.
\nonumber\\
&&+ \left.\left[\frac13( 1-a_1+3d_1) r^3-\frac12(-r_0+b_1+c_1)r^2 - \frac12( r_0^2+b_2+c_2)r 
	+\frac13(-r_0^3-a_2+ 3d_2)\right] \right\}\nonumber\\
 &&+ O(\theta^2), \quad r<r_0.
\eea\label{psiat0}\esube
Similarly, for $\theta$ near $\pi$, we have 
\bea
\Psi &=& \tilde A\frac23\left\{\phantom{\frac12}\hspace{-4mm}
 \left[(-1+a_1+3d_1) r^3 + 3(-r_0+b_1-c_1)r^2 + 3(-r_0^2+b_2-c_2)r -r_0^3+a_2+3d_2\right] 
	(\theta-\pi)^{-2}\right.
\nonumber\\
&&+ \left.\left[\frac13(-1+a_1+3d_1) r^3-\frac12(-r_0+b_1-c_1)r^2 - \frac12(-r_0^2+b_2-c_2)r 
	+\frac13(-r_0^3+a_2+ 3d_2)\right] \right\}\nonumber\\
 &&+ O[(\theta-\pi)^2].
\eea

We thus obtain as necessary and sufficient conditions for smoothness of $\Psi$ at $\theta=0$,
\bsube\bea
a_1-3d_1 &=&-1, \quad a_2-3d_2 =  r_0^3,\quad b_1+c_1 =-r_0,\quad b_2+c_2 =  r_0^2, 
\quad r>r_0; \\
a_1-3d_1 &=&  1, \quad a_2-3d_2 =-r_0^3,\quad b_1+c_1 = r_0,\quad b_2+c_2 = -r_0^2, 
\quad r<r_0;
\eea\esube
and at $\theta = \pi$, 
\be
a_1+3d_1 =  1, \quad a_2+3d_2 =  r_0^3,\quad b_1-c_1 =  r_0,\quad b_2-c_2 =  r_0^2.
\ee

Because the perturbations corresponding to $d_1$ and $d_2$ vanish, one can
satisfy the first two conditions in each line without changing $h_{\alpha\beta}$.
The necessary and sufficient 
conditions for smoothness of $h_{\alpha\beta}$ at $\theta=0$ (as can be verified 
directly from its components) are then 
\bsube\bea
 b_1+c_1 &=& -r_0, \quad  b_2+c_2 =  r_0^2, \quad r>r_0;\label{smooth0>} \\
 b_1+c_1 &=&  r_0, \quad  b_2+c_2 = -r_0^2, \quad r<r_0;\label{smooth0<}
\eea\label{smooth0}\esube
and at $\theta=\pi$,
\be
 b_1-c_1 =  r_0, \quad b_2-c_2 =  r_0^2. \\
\label{smoothpi}\ee

For the two cases, these parameters -- 
the unique solutions to Eqs.~(\ref{smoothpi}) and to either Eqs.~(\ref{smooth0>}) or  
(\ref{smooth0<}) -- are listed in Table \ref{fs_regularity}.

As in the last section, Eqs.~(\ref{smooth0}) immediately imply that one cannot find a radiation gauge that can be everywhere locally obtained from a Hertz potential, for which $h_{\alpha\beta}$ is simultaneously smooth at $\theta = 0$ outside and inside the particle.  
One can choose parameters for which $h_{\alpha\beta}$ smooth everywhere except \\
(I) along the part of the $\theta=0$ ray below the particle (with $r\leq r_0$), or\\
(II) along the part of the $\theta=0$ ray above the particle (with $r\geq r_0$).
For the parameter choice that makes $h_{\alpha\beta}$ smooth outside $r=r_0$, however, 
$b_2$ has the nonzero value $-r_0^2$, changing the asymptotic mass 
to zero by subtracting a mass $\mathfrak m$ Schwarzschild solution centered at the origin. 
As in the Schwarzschild perturbation of the last section, one can then obtain a smooth 
metric outside $r=r_0$ with asymptotic mass $\mathfrak m$ by adding back an $l=0$ perturbation 
in a smooth gauge -- adding, for example the linearized Schwarzschild solution of mass 
$\mathfrak m$ centered at the origin.     
 
\begin{table}[h]
\caption{Singularity location for two choices of parameters}
\label{fs_regularity}
\begin{center}
\begin{tabular}{| c | c | c |}
\hline
Parameter  & (I) Singularity at &(II) Singularity at\\
 & $\theta=0$; $r\geq r_0$ & $\theta=0$; $r\leq r_0$\\
\hline
$a_1$ & $0$ & $0$\\
$a_2$ & 0& 0\\
$b_1$ & $\!\! r_0$ & $0$\\
$b_2$  & $0$ & $\!\! r_0^2$\\
$c_1$  &$0$ & $-r_0$ \\
$c_2$  & $-r_0^2$ & $0$ \\
\hline
\end{tabular}
\end{center}
\end{table}

\noindent

The perturbed metric smooth everywhere except $\theta=0,\ r\geq r_0$ is given by
\bsube\label{hfs1}\begin{eqnarray}
h_{22}^{(I)}&=&-2{\mathfrak m}\left(\frac{1}{r_0}-\frac{1}{R}\right),\\
h_{23}^{(I)}&=& -\sqrt{2}\,{\mathfrak m}\frac1{\sin\theta}\left(\frac{r_0}{r R}
	 -\frac{1}{r}-\frac1R\cos\theta\right),\\
h_{33}^{(I)}&=&-2{\mathfrak m}\frac1{\sin^2\theta}
\left[\frac{1}{R}-\frac{1}{r_0}+2
   \left(\frac{1}{r}-\frac{r_0 }{r R}-\frac{r}{r_0 R}\right)\cos\theta
  +\left(\frac{3}{R}-\frac{1}{r_0}\right)\cos^2\theta\right].
\end{eqnarray}\esube

The perturbed metric smooth everywhere except  $\theta=0,\ r\leq r_0$ is given by
\bsube\label{hfs2}\begin{eqnarray}
h_{22}^{(II)}&=&-2{\mathfrak m}\left(\frac{1}{r}-\frac{1}{R}\right),\\
h_{23}^{(II)}&=&- \sqrt{2}\ {\mathfrak m}\frac{1}{\sin\theta}
	\left[\frac{r_0}{rR}+\left(\frac1r-\frac1R\right)\cos\theta 
	 \right],\\
h_{33}^{(II)}&=&-2{\mathfrak m}\frac{1}{\sin^2\theta}
\left[\frac1R-\frac1r 
	+2\left(\frac1{r_0}-\frac r{r_0R}-\frac{r_0}{rR}\right)\cos\theta 
	+ \left(\frac3R-\frac1r\right)\cos^2\theta
\right].
\end{eqnarray}\esube

The perturbed metrics $h^{(I)}_{\alpha\beta}$ and 
$h^{(II)}_{\alpha\beta}+h_{\alpha\beta}^{\rm Schwarzschild}({\mathfrak m})$ must 
agree, up to a gauge transformation, 
with the perturbation one acquires from the Schwarzschild metric by translating 
the origin and linearizing about flat space.  This is the case, and the explicit 
gauge vector $\xi^\alpha$ for the metric $h^{I}_{\alpha\beta}$ has components 
\bsube\label{fs_scw_gauge}\begin{eqnarray}
\xi_1&=&-{\mathfrak m}\left(\frac{t-r}{r_0}-\frac{r-r_0\cos\theta}{R}+2\ln\frac{r - r_0\cos\theta+R}{\sin^2\frac \theta 2}\right),\\
\xi_2&=&-\frac {\mathfrak m} 2\left(\frac{t-3r}{r_0}+\frac{r-r_0\cos\theta}{R}+2\ln\frac{r - r_0\cos\theta+R}{\sin^2\frac \theta 2}\right),\\
\xi_3&=&-\frac{{\mathfrak m}}{2\sqrt{2}\sin\theta}\left\{\frac{4r}{r_0}\cos\theta+\frac{2[R^2+(r-r_0\cos\theta)^2]}{r_0 R}-4\right\}.
\end{eqnarray}\esube

The resultant metric is given by
\bsube\begin{eqnarray}
h_{11}&=& \frac{2\mathfrak m}{R^3}(2R^2-r_0^2\sin^2\theta),\\
h_{22}&=& \frac{\mathfrak m}{2R^3}(2R^2-r_0^2\sin^2\theta),\\
h_{33}&=& \frac{\mathfrak m r_0^2\sin^2\theta}{R^3},\\
h_{12}&=& \frac{\mathfrak m r_0^2\sin^2\theta}{R^3},\\
h_{13}&=& \frac{\sqrt 2\mathfrak m r_0(r-r_0\cos\theta)\sin\theta}{R^3},\\
h_{23}&=& \frac{\mathfrak m r_0(r-r_0\cos\theta)\sin\theta}{\sqrt 2R^3},\\
h_{34}&=& \frac{\mathfrak m r_0^2\sin^2\theta}{R^3}.
\end{eqnarray}\esube

\section{Finding the self-force in a radiation gauge.}
\label{renormalization}

An algorithm for finding the self force on a mass moving in a background 
spacetime, the MiSaTaQuWa method, was given several years ago 
\cite{MinoSasaki,QuinnWald}.  In this 
method, as noted by Quinn and Wald and by Detweiler and Whiting 
\cite{DetWhiting}, a particle follows a geodesic of a renormalized metric, 
$h^{\rm ren}$, given by 
\be 
 h^{\rm ren} = h^{\rm ret} - h^{\rm sing},
\ee
where $h^{\rm ret}$ is the retarded field of the particle in a Lorenz 
gauge and 
$h^{\rm sing}$ is its locally defined singular part, chosen to cancel the 
singular part of $h^{\rm ret}$ and to give no contribution to the self-force.
In the Quinn-Wald 
characterization of the method $h^{\rm sing}$ is obtained in a neighborhood 
of a point $P$ of the particle's trajectory from the 
field of a particle moving in flat space by using the exponential map 
to relate the linearized field in flat space to a field $h^{\rm sing}$ 
in the neighborhood of $P$.  One demands that the image under the 
exponential map of the flat-space particle's position, 4-velocity, and 
acceleration agree at $P$ with those of the original particle.

Because it involves a Lorenz gauge, the method cannot take advantage of 
the decoupled, separable wave equation -- the Teukolsky equation -- that 
governs black hole perturbations.  One could imagine constructing the 
singular part of the gauge-invariant perturbed Weyl tensor component 
$\Psi_0$ (or $\Psi_4$) from $h^{\rm sing}$ and then obtaining the renormalized 
metric in a radiation gauge from $\Psi_0^{\rm ren}$ (or $\Psi_4^{\rm ren}$).
The radiation-gauge prescription, however, requires the perturbed Weyl scalar 
to be a solution to the linearized Einstein equation. In the forms in which it 
is given by Mino et al.\cite{MinoSasaki} and by Quinn and Wald\cite{QuinnWald},  $h^{\rm sing}$ is not 
a solution to the Einstein equations linearized about the background 
Kerr spacetime; and it is not obvious that one can obtain a prescription 
for a renormalized Weyl scalar $\Psi_0^{\rm ren}$ (or $\Psi_4^{\rm ren}$) 
that satisfies the Teukolsky equation.    

A more recent version of the MiSaTaQuWa method, due to Detweiler and Whiting
\cite{DetWhiting}, 
overcomes this difficulty.
In their prescription, the singular part of the field $h^{\rm sing}$ is redefined 
in a way that makes it a solution to the linearized Einstein equations, defined in 
a normal neighborhood of the particle.  As a 
result, one can compute a gauge-invariant renormalized Weyl scalar in the 
following way: \\
The perturbed Riemann tensor is given in terms of $h_{\alpha\beta}$ by 
\be
\dot R_{\alpha\beta\gamma\delta}
= \frac12 (\nabla_\beta\nabla_\gamma h_{\alpha\delta}
	+ \nabla_\alpha\nabla_\delta h_{\beta\gamma}
	- \nabla_\alpha\nabla_\gamma h_{\beta\delta}
	-\nabla_\beta\nabla_\delta h_{\alpha\gamma})
	-R_{\alpha\beta[\gamma}{}^\epsilon h_{\delta]\epsilon}.
\ee

From Eq. (\ref{Psi_i}), $\psi_0$ has the form,
\be 
   \psi_0 = -l^\alpha m^\beta l^\gamma m^\delta \dot C_{\alpha\beta\gamma\delta} 
	= -l^\alpha m^\beta l^\gamma m^\delta \dot R_{\alpha\beta\gamma\delta}
	= {\cal O}^{\alpha\beta}h_{\alpha\beta},
\ee 
where ${\cal O}^{\alpha\beta}$ is the operator 
\be 
{\cal O}^{\alpha\beta} 
	= \frac12(l^\alpha l^\beta m^\gamma m^\delta +m^\alpha m^\beta l^\gamma l^\delta
	-l^\alpha m^\beta l^\gamma m^\delta-l^\alpha m^\beta m^\gamma l^\delta)
\nabla_\gamma\nabla_\delta.
\ee 
In particular, from the singular part of the perturbed metric, $h^{\rm sing}_{\alpha\beta}$,
one can define the singular part $\psi^{\rm sing}_0$ of the perturbed Weyl tensor  by 
\be 
\psi^{\rm sing}_0 \equiv {\cal O}^{\alpha\beta}h_{\alpha\beta}^{\rm sing}.
\ee

Then $\psi_0^{\rm ren}$ is given by
\be
\psi_0^{\rm ren} = \psi_0^{ret} - \psi_0^{\rm sing},
\ee 
with 
\be
\psi_0^{\rm sing} = {\cal O}^{\alpha\beta}h^{\rm sing}_{\alpha\beta}.
\ee 
Because $h^{\rm ret}$ and $h^{\rm sing}$ have the same source, so do $\psi^{\rm ret}$ 
and $\psi^{\rm sing}$.   Their difference, $\psi^{\rm ren}$ is thus a {\em sourcefree} 
solution to the Teukolsky equation.  That implies that one can construct 
a nonsingular $h^{\rm ren}$ in a radiation gauge from $\Psi^{\rm ren}$.

In a Lorenz gauge, we have    
\be
\bar h^{\rm sing}_{\alpha\beta} 
 = 4 \int G^{{\rm sing}}_{\alpha\beta\gamma'\delta'}(x,x')T^{\gamma'\delta'}(x') d^4V',
\ee
where \dis \bar h^{\alpha\beta}= h^{\alpha\beta}-\frac12 g^{\alpha\beta} h$ and 
where the Green's function has the form  
\be
  G^{{\rm sing}}_{\alpha\beta\gamma'\delta'}(x,x') 
	= \frac12 U_{\alpha\beta\gamma'\delta'}(x,x')\delta(\sigma)
	+ \frac12 V_{\alpha\beta\gamma'\delta'}(x,x')\Theta(\sigma).
\ee
Here $U$ and $V$ are nonsingular bitensors, whose form is given, for 
example, in \cite{DetWhiting} (see also \cite{Poisson}).
The step function $\Theta(\sigma)$ has value $1$ when $\sigma>0$ -- when 
$x$ and $x'$ are {\em spacelike} separated; and it vanishes for $\sigma<0$.  
Thus $G^{{\rm sing}}_{\alpha\beta\gamma'\delta'}(x,x')$ has peculiar causal behavior: 
It is nonzero only when $x$ and $x'$ are not timelike related.      

A Green's function for $\Psi_0^{\rm sing}$ is then given by  
\be 
  G^{\rm sing}_{\gamma'\delta'}(x,x') 
  = \left({\cal O}^{\alpha\beta}-\frac12 g^{\alpha\beta}{\cal O}^\epsilon_\epsilon\right)
	G^{\rm sing}_{\alpha\beta\gamma'\delta'}(x,x').
\ee
Having computed $G^{\rm sing}_{\gamma'\delta'}(x,x')$, one can obtain 
$\psi_0^{\rm sing}$ from the 
expression
\be 
   \psi_0^{\rm sing}(x) 
	= \int G_{\gamma'\delta'}(x,x')T^{\gamma'\delta'}(x')d^4V'.
\ee
The explicit forms of $U$ and $V$ have been computed in terms of the 
geodesic distance from the particle, to the order needed to find the self-force
(although the order needed for the prescription given here 
may be different).   

From $\psi_0^{\rm ren}$ one recovers $h^{\rm ren}_{\alpha\beta}$, using the 
CCK equations (\ref{gr_scw_psi0}) or (\ref{gr_scw_psi4}) in a Kerr background to find the potential $\Psi$
and using Eq.~(\ref{kc_metric}) to find $h^{\rm ren}$ in terms of $\Psi$.  
Eq.~(\ref{gr_scw_psi0}) can be solved by four radial integrations 
for each angular harmonic if one works in the frequency domain. 
The operator $D$, restricted to an $m,\omega$ subspace corresponding 
to $\phi$ and $t$ dependence \dis e^{i(m\phi-\omega t)}$, has the form
\be 
D_{m\omega}= \partial_r + i\frac {[ma-(r^2+a^2)\omega]}\Delta .
\ee
Then $\Psi^{\rm ren}$ is found from 
\be 
\psi_{0,\ m\omega}^{\rm ren}=D_{m\omega}^4\Psi_{m\omega}^{\rm ren}.
\label{Psi_m}
\ee
Because $h^{\rm ren}_{\alpha\beta}$ is a solution to the linearized vacuum 
Einstein equations, it is determined by $\psi_0$ (or $\psi_4$) up to an 
algebraically special vacuum perturbation. \\   

The method of finding the self-force in a radiation gauge 
may now be summarized as follows: 
\begin{enumerate}
\item Compute $\psi_0^{\rm sing}$ either from the known form of 
$h^{\rm sing}$ as outlined above, or, with additional insight, 
directly from the Teukolsky equation.
\item Express $ \psi_0^{\rm sing}$ in terms of spin-weighted spheroidal 
harmonics.  
\item Compute $\psi_0^{\rm ret}$ from the Teukolsky equation.
\item Write $\psi_0^{\rm ren}=  \psi_0^{\rm ret} - \psi_0^{\rm sing}$, 
regularizing by a cutoff in angular harmonics, and finding the 
limit, to desired accuracy, as the cutoff harmonic is increased.
\item Find the potential $\Psi$ either from $\psi_4^{\rm ren}$, using 
Eq.~(\ref{gr_scw_psi4}), or from $\psi_0^{\rm ren}$ by 4 radial integrations  of 
Eq.~(\ref{Psi_m}).
\item Find the perturbed metric $h_{\alpha\beta}$ in a radiation 
gauge by taking derivatives of $\Psi$. This leaves spin-weight 0 and 1 
parts of the perturbation undetermined.
\item Obtain the spin-weight 0 and 1 parts of $h_{\alpha\beta}$ by 
fixing the area and angular velocity of the perturbed black hole and 
using jump conditions across the particle of the spin-weight 0 and 
1 parts of the perturbed Einstein equation.\footnote{Using the 
jump conditions of the spin-weight 0 and 1 parts of the Einstein 
equation was suggested to us by Bernard Whiting.}
\item Compute the self-force from the perturbed geodesic equation.
For the perturbed metric $g_{\alpha\beta}+\zeta\,h_{\alpha\beta}$, 
with $\nabla$ the covariant derivative of $g_{\alpha\beta}$ 
and $u^\alpha$ normalized by $g_{\alpha\beta}$, the geodesic 
equation has, to $O(\zeta)$, the form 
\be 
u\cdot \nabla u^\alpha  
= -\zeta(g^{\alpha\delta}+u^\alpha u^\delta)
	\left(\nabla_\beta h^{\rm ren}_{\gamma\delta}
	-\frac12\nabla_\delta h^{\rm ren}_{\beta\gamma}\right)\equiv f^\alpha.
\label{geopert}\ee 
\end{enumerate}

With the self-force $f^\alpha$ computed, one must find the particle trajectory as a self-consistent solution to Eq.~(\ref{geopert}). The emitted radiation can 
be found from $\psi_4^{\rm ret}$.   

Because the perturbed Weyl tensor involves two derivatives of the perturbed metric,
it is possible that the renormalization program outlined here will require 
a Hadamard expansion that is two orders higher in the separation between 
source and field point than the expansion used in a Lorenz gauge.  We suspect, 
however, that the extra orders can be avoided by using Eq.~(\ref{Psi_m}) to find the 
Hertz potential.  The argument is that each radial integration changes by one 
factor of $R$ the highest power of $1/R$ and thereby reduces the order 
of the singularity.  Calculations now underway in a Schwarzschild background 
will decide the issue and may show whether the method 
can feasibly be used to obtain generic orbits in a Kerr background.

\acknowledgments  We thank Steven Detweiler, Eric Poisson, Robert Wald, Bernard Whiting, and participants in the Capra 6 meeting for helpful conversations, the referee for the detailed review and comments on this paper, and Ishai Ben-Dov for access to unpublished notes. 
This work was supported in part by NSF Grants. PHY 0503366 and PHY 0200852.  T. S. Keidl is supported by NASA Wisconsin Space Grant Consortium.

\appendix
\section {Smoothness in terms of a Hertz potential}
\label{appendixa}

In the CCK formalism one writes the vector potential $A_\alpha$ and 
the perturbed metric $h_{\alpha\beta}$ in 
terms of Hertz potentials $\Phi$ and $\Psi$.  The components of 
$A_\alpha$ and $h_{\alpha\beta}$, given by Eqs.~(\ref{a_scw_irg}) and 
(\ref{gr_scw_irg}),
have in each case the form ${\cal L}\Phi$ (or ${\cal L}\Psi$), where 
$\cal L$ is a sum of covariant derivatives along the basis vectors 
and Christoffel symbols (spin coefficients) of the basis.  Because 
the Christoffel symbols of a smooth basis are smooth, smoothness of 
the Hertz potential $\Phi^\pm$ associated with a smooth basis $\{\bf e_\mu^\pm\}$ is a sufficient condition for smoothness of $A_\alpha$; and smoothness of the Hertz potential $\Psi^\pm$ associated with a smooth basis is a sufficient condition for smoothness of $h_{\alpha\beta}$.  

We next relate this condition to a condition on the Hertz potential associated 
with the Kinnersley basis, by showing that the Hertz potentials $\Phi^\pm$ and 
$\Psi^\pm$ associated with the smooth bases $\{\bf e_\mu^\pm\}$ differ 
only by a phase from the Hertz potentials $\Phi$ and $\Psi$.  In addition, 
the metrics corresponding to the two sets of Hertz potentials are 
identical.

These results are immediate from the following proposition, proved here 
for a Schwarzschild geometry, but presumably correct for type D vacuum spacetimes.

\noindent
{\em Proposition}. (i) Let $\Psi$ satisfy the sourcefree Bardeen-Press
(Teukolsky) equation for spin-weight $s=2$, namely ${\cal T}\bar\Psi = 0$.  Then 
$\Psi^\pm = \Psi e^{\mp 2i\phi}$  satisfies the same sourcefree equation, 
${\cal T}\bar\Psi^\pm = 0$, and 
$h_{\alpha\beta}[\Psi^\pm]=h_{\alpha\beta}[\Psi]$.
\\
(ii) Let $\Phi$ satisfy the sourcefree Teukolsky  
equation for spin-weight $s=1$, ${\cal T}\bar\Phi = 0$.  Then 
$\Phi^\pm = \Phi e^{\mp i\phi}$  satisfies the same sourcefree equation, 
${\cal T}\bar\Phi^\pm = 0$, and $A_{\alpha}[\Phi^\pm]=A_{\alpha}[\Phi]$.\\

\noindent
{\em Proof}. The proof essentially follows from the fact that expressions 
arising in the NP and CCK formalism have definite spin-weight.  A Lorentz 
tranformation of the basis that changes only the phase of $m^\alpha$  changes
the phase of each expression only by a phase corresponding to the spin 
weight of that expression.  The detailed verification is as follows.
The operators and spin coefficients associated with the bases $\{\bf e_\mu^\pm\}$ 
are related to the corresponding objects of the Kinnersley basis by 
\bsube\bea 
D^\pm &=& D, \quad {\bm \Delta}^\pm = {\bm \Delta}, \quad 
\delta^\pm = e^{\pm i\phi}\delta, \quad 
\overline\delta^\pm = e^{\mp i\phi}\overline\delta\\
\beta^\pm
 &=& -\alpha^\pm=\beta - \frac1{2\sqrt2\, r\sin\theta}e^{i\phi},\quad
\gamma^\pm=\gamma, \quad \rho^\pm=\rho,\quad \mu^\pm=\mu.
\eea
\label{pm}
\esube
For a quantity $\eta$ having spin-weight 2, we have 
$\eta^\pm=e^{\pm2i\phi}\eta$, 
and from Eqs.~(\ref{pm}) above we obtain the relations
\bsube\bea
(D+n\rho)^\pm\eta^\pm 
&=& e^{\pm2i\phi}(D+n\rho)\eta,\quad \mbox{any } n,
\label{pmopa}\\
\left({\bm \Delta}+2\gamma+\mu\right)^\pm \eta^\pm 
&=& e^{\pm2i\phi}\left({\bm \Delta}+2\gamma+\mu\right)\eta,
\label{pmopc}\\ 
(\overline\delta+4\beta)^\pm\eta^\pm
&=& e^{\pm i\phi}(\overline\delta+4\beta)\eta,
\label{pmopd}
\eea\label{pmop}\esube
Similarly, a quantity $\eta$ of spin-weight 1 satisfies the relations
\bsube\bea
D^\pm\eta^\pm &=& e^{\pm i\phi}D\eta,
\label{pmop2a}\\
(\overline\delta+2\beta)^\pm\eta^\pm
&=& (\overline\delta+2\beta)\eta, 
\label{pmop2b}\\
(\delta-2\beta)^\pm\eta^\pm
&=& e^{\pm 2i\phi}(\delta-2\beta)\eta.
\label{pmob2c}\eea\label{pmop2}\esube

From the form of the Teukolsky spin-2 and spin-1 operators of Eqs.~(\ref{gr_scw_scalar_eqn}) 
and (\ref{em_ck_scalar}), 
\bea
{\cal T}_{s=2} &=& ({\bm \Delta}+2\gamma+\mu)(D+3\rho)
	-(\delta -2\beta)(\overline\delta+4\beta)-3\Psi_2,\\
{\cal T}_{s=1}&=& ({\bm \Delta}+\mu)(D+\rho)
	-\delta(\overline\delta+2\beta),
\eea
and Eqs.~(\ref{pmop}) and (\ref{pmop2}), we have 
the claimed relations  ${\cal T}^\pm\Psi^\pm = {\cal  T}\Psi$, 
$\ {\cal T}^\pm\Phi^\pm = {\cal  T}\Phi$.

From the form (\ref{gr_scw_irg}) of $h_{\alpha\beta}$, Eqs.~(\ref{pmop}) 
imply that the components $h_{\alpha\beta}[\Psi^\pm]$ along the basis 
$\{\bf e^\pm_\mu\}$ are related to the components of $h_{\alpha\beta}[\Psi]$ 
along the Kinnersley basis by  
\be
h_{22}[\Psi^\pm] = h_{22}[\Psi], 
\quad h_{23}[\Psi^\pm] = e^{\pm i\phi} h_{23}[\Psi], \quad
h_{33}[\Psi^\pm] = e^{\pm 2i\phi} h_{33}[\Psi].
\ee
But these are just the components of $h_{\alpha\beta}[\Psi]$ in the 
basis $\{\bf e^\pm_\mu\}$, implying 
$h_{\alpha\beta}[\Psi^\pm]=h_{\alpha\beta}[\Psi]$, 
as claimed.

Similarly, from the form (\ref{a_scw_irg}) for $A_\alpha$, Eqs.~(\ref{pmop2})
imply $A_\alpha[\Phi^\pm] = A_\alpha[\Phi]$. $\Box$ 

\section{$\phi_2$ and $\psi_4$ in closed form for static particles.}
\label{appendixb}

We first show that the expression~(\ref{em_phi2_mode}) for $\phi_2$ as a series has 
the sum (\ref{em_phi2_closed}) and then turn to the gravitational case, showing  
that the series~(\ref{psi4_mode}) has the sum (\ref{gr_psi4_closed}).  

Noting that the sum (\ref{em_phi2_mode}) is  
axisymmetric and using Eq.~(\ref{green_sylm_prop_3}), we can write the sum in the form,  
\begin{eqnarray}
\phi_2&=&\frac{4\pi e (\Delta
\Delta_0)^{1/2}}{2\sqrt{2}Mr^2r_0}\sum_{l=1}^\infty\frac{1}{[l(l+1)]^{1/2}}P^1_l\left(\frac{{\mathfrak
r}_<}{M}\right)Q^1_l\left(\frac{{\mathfrak r}_>}{M}\right)
        {}_{-1}\!Y_{l 0}(\theta,\phi)Y^*_{l 0}(0,0)\\
&=&-\frac{4\pi e (\Delta
\Delta_0)^{1/2}}{2\sqrt{2}Mr^2r_0}\bar\eth\sum_{l=1}^\infty\frac{1}{l(l+1)}P^1_l\left(\frac{{\mathfrak
r}_<}{M}\right)Q^1_l\left(\frac{{\mathfrak r}_>}{M}\right)
        \!Y_{l 0}(\theta,\phi)Y^*_{l0}(0,0)\\
&=&-\frac{ e (\Delta
\Delta_0)^{1/2}}{2\sqrt{2}Mr^2r_0}\bar\eth\sum_{l=1}^\infty\frac{(2l+1)}{l(l+1)}P^1_l\left(\frac{{\mathfrak
r}_<}{M}\right)Q^1_l\left(\frac{{\mathfrak r}_>}{M}\right)\!P_l(\cos\theta).
\label{emsumpl}
\end{eqnarray}

To sum the series (\ref{emsumpl}), we will use the relation
\be
S \equiv \frac 1 {\sqrt{a^2+b^2-2abx-(1-x^2)}} = \sum_{l=0}^\infty (2l+1) P_l (a) Q_l(b) P_l(x), 
\qquad b\ge a\ge 1,
\label{S}
\ee
which follows immediately from the series expression for the Laplacian Green's 
function, $1/R$, written in prolate spheroidal coordinates \cite{MacRobert}. 
That is, the distance $R$ between points with prolate spheroidal coordinates $(r,\theta,\phi)$ and 
$(r_1,\theta_1,\phi_1)$ is given by
\be
R^2=r^2+r_1^2-2r r_1\cos\theta\cos\theta_1-2\sqrt{r^2-\varepsilon^2}\sqrt{r_1^2-\varepsilon^2}\sin\theta\sin\theta_1\cos(\phi-\phi_1)
-\varepsilon^2(\sin^2\theta+\sin^2\theta_1),
\label{rprolate}\ee 
where $r_1,r>\varepsilon$; and the Green's function is given for $r_1>r$ by the series   
\begin{eqnarray}
\frac 1 R\! & = &\!\! \frac 1 \varepsilon \bigg[\sum_{l=0}^\infty (2l+1)[P_l\left(\frac r \varepsilon \right)Q_l \left(\frac {r_1} \varepsilon \right) P_l(\cos\theta)P_l(\cos\theta_1)+ \nonumber\\
& &\!\! 2\sum_{m=1}^\infty(-1)^m\!\!\left\{\frac{(l-m)!}{(l+m)!}\right\}^2\!\!
P_l^m\!\!\left(\frac r \varepsilon \right)Q_l^m \!\!\left(\frac {r_1} \varepsilon \right)P_l^m(\cos\theta)P_l^m(\cos\theta_1)\cos m(\phi-\phi_1)]\bigg].
\label{gprolate}\end{eqnarray}
Setting $\theta_1=0$,  
\dis\ \frac {r} \varepsilon=a$,  \dis\ \frac {r_1} \varepsilon=b$, and $x=\cos\theta$ in Eqs.~(\ref{rprolate}) and (\ref{gprolate}), we obtain Eq.~(\ref{S}).

We next use the standard identities \cite{Gradstein},
\bsube
\bea
(2l+1) x \left[{P_l(x)}\atop{Q_l(x)}\right] 
= (l+1) \left[{P_{l+1}(x)}\atop{Q_{l+1}(x)}\right]
+l \left[{P_{l-1}(x)}\atop{Q_{l-1}(x)}\right]\label{id_xPl},
\label{idxPl}\\
(x^2-1)\frac d{dx} \left[{P_l(x)}\atop{Q_l(x)}\right]
=(l+1)\left\{\left[{P_{l+1}(x)}\atop{Q_{l+1}(x)}\right]
-x \left[{P_l(x)}\atop{Q_l(x)}\right]\right\}
\label{id_P_der},
\eea\label{idPl}\esube
to show the relation
\bsube\bea
S_1&\equiv&(-ab +x)S+a \label{S1a}\\
&=&(a^2-1)^{1/2}(b^2-1)^{1/2}\sum_{l=1}^\infty \frac{(2l+1)}{l(l+1)} P_l^1(a)Q_l^1(b)P_l(x).\label{S1}
\eea\esube
First, from the definition (\ref{S}) of $S$ and Eq.~(\ref{id_xPl}), we obtain
\bea
x S & = & \sum_{l=0}^\infty \left[ l P_{l-1}(a)Q_{l-1}(b)+(l+1)P_{l+1}(a)Q_{l+1}(b) \right] P_l(x).\label{xS}
\eea

Now, substituting Eq.~(\ref{xS}) in Eq.~(\ref{S1a}), we have
\bea
S_1
&=&-\sum_{l=0}^\infty[(2l+1)aP_l(a)bQ_l(b)-l P_{l-1}(a)Q_{l-1}(b)
-(l+1)P_{l+1}(a)Q_{l+1}(b)]P_l(x)+ a\nonumber\\
&=&-\sum_{l=1}^\infty[(2l+1)aP_l(a)bQ_l(b)-l P_{l-1}(a)Q_{l-1}(b)-(l+1)P_{l+1}(a)Q_{l+1}(b)]P_l(x),
\eea
where the relations $P_0=1,\ P_1(a)=a,\ Q_1(b) =bQ_0(b)-1$, are used to eliminate the $l=0$ terms. 
Again from Eq.~(\ref{id_xPl}), we obtain
\bea
S_1&=&-\sum_{l=1}^\infty\left\{(2l+1)
\left[\frac{(l+1)}{(2l+1)}P_{l+1}(a)+\frac l {(2l+1)} P_{l-1}(a)\right]
\left[\frac{(l+1)}{(2l+1)}Q_{l+1}(b)+\frac l {(2l+1)}Q_{l-1}(b)\right]\right.\nonumber\\
& &\left.\phantom{xxx}\qquad-l P_{l-1}(a)Q_{l-1}(b)-(l+1)P_{l+1}(a)Q_{l+1}(b)
\phantom{\frac AB}\!\!\!\!\!\!\right\}P_l(x)
\nonumber\\
&=&\sum_{l=1}^\infty\frac{l(l+1)}{2l+1}\left[P_{l+1}(a)Q_{l+1}(b)-P_{l-1}(a)Q_{l+1}(b)-P_{l+1}(a)Q_{l-1}(b)+P_{l-1}(a)Q_{l-1}(b)\right]P_l(x)\nonumber\\
&=&\sum_{l=1}^\infty\frac{l(l+1)}{2l+1}\left[P_{l+1}(a)-P_{l-1}(a)\right]\left[Q_{l+1}(b)-Q_{l-1}(b)\right]P_l(x).
\label{S1c}
\eea

Using the definitions,
\be
P_l^m(x)=(x^2-1)^{m/2} \frac {d^m}{dx^m} P_l(x),\quad Q_l^m(x)=(x^2-1)^{m/2} \frac {d^m}{dx^m} Q_l(x), \quad |x|\geq 1,
\label{Plm}\ee
and the identities~(\ref{id_P_der}) and~(\ref{id_xPl}), we can write
\bea
P_l^1(a)&=&\frac{l(l+1)}{(2l+1)(a^2-1)^{1/2}} \left[P_{l+1}(a)-P_{l-1}(a)\right]
\label{Pl1}.
\eea
Using this relation and the analogous relation for $Q_l^1(b)$ to replace the 
bracketed expressions in Eq.~(\ref{S1c}) yields Eq.~(\ref{S1}), as claimed.

Finally, from Eqs.~(\ref{emsumpl}) and (\ref{S1}), with \dis a=\frac{{\mathfrak r}_<}M, 
b=\frac{{\mathfrak r}_>}M,\ x=\cos\theta$,
we obtain the closed form expression (\ref{em_phi2_closed}) for $\phi_2$: 
\begin{eqnarray}
\phi_2&=&\frac{ e (\Delta \Delta_0)^{1/2}}{2\sqrt{2}Mr^2r_0}\bar\eth
\left[\frac M {(\Delta\Delta_0)^{1/2}}\frac{\mathfrak r \mathfrak
r_0-M^2\cos\theta}{R}\right]\\
&=&e \frac{\Delta_0}{2 \sqrt2\,r_0}  \frac{\Delta\sin\theta}{r^2R^3}.
\end{eqnarray}

Next, we turn to the gravitational series.  Again we use axisymmetry and Eq.~(\ref{green_sylm_prop_3}) to write the sum (\ref{psi4_mode}) in the form
\bea
\psi_4&=&-\frac{\pi {\mathfrak m}}{M} \frac{\Delta\Delta_0^{1/2}}{r^4
r_0}\sum_{l=2}^\infty 
\left[\frac{(l-2)!}{(l+2)!}\right]^{1/2}P_l^2\left(\frac{{\mathfrak
r}_<}{M}\right) Q_l^2\left(\frac{{\mathfrak r}_>}{M}\right)
{}_{-2}\!Y_{l 0}(\theta,\phi) Y_{l 0}^*(0,0)
\nonumber\\
&=&-\frac{\pi {\mathfrak m}}{M} \frac{\Delta\Delta_0^{1/2}}{r^4
r_0}\bar\eth^2\sum_{l=2}^\infty \frac{(l-2)!}{(l+2)!}
P_l^2\left(\frac{{\mathfrak r}_<}{M}\right) Q_l^2\left(\frac{{\mathfrak
r}_>}{M}\right) Y_{l 0}(\theta,\phi) Y_{l 0}^*(0,0)\nonumber\\
&=&-\frac{{\mathfrak m}}{4 M} \frac{\Delta\Delta_0^{1/2}}{r^4
r_0}\bar\eth^2\sum_{l=2}^\infty (2l+1)\frac{(l-2)!}{(l+2)!}
P_l^2\left(\frac{{\mathfrak r}_<}{M}\right) Q_l^2\left(\frac{{\mathfrak
r}_>}{M}\right) P_l(\cos\theta).
\label{grsumpl}
\eea

In this case, we wish to show
\bea
S_2&\equiv&(a^2+b^2+a^2 b^2-1-4abx+2x^2)S-(a^2+1)b-a(a^2-3)x\nonumber\\
&=&(a^2-1)(b^2-1)\sum_{l=1}^\infty \frac{(2l+1)}{(l+2)(l+1)l(l-1)} P_l^2(a)Q_l^2(b)P_l(x).\label{S2}
\eea

Using Eq.~(\ref{xS}), Eq.~(\ref{id_xPl}), and relabeling the indices, we obtain
\bea
x^2 S & = & \sum_{l=0}^\infty \left\{ \frac{l(l-1)}{(2l-1)} P_{l-2}(a)Q_{l-2}(b)+\left[\frac{(l+1)^2}{(2l+3)}+\frac{l^2}{(2l-1)}\right]P_l(a)Q_l(b)\right.\nonumber\\
& &+\left.\frac{(l+1)(l+2)}{(2l+3)} P_{l+2}(a)Q_{l+2}(b)\right\} P_l(x).\label{x2S}
\eea
Similarly, from Eq.~(\ref{idxPl}), we have
\bea
a^2P_l(a)&=&\frac{(l+1)(l+2)}{(2l+1)(2l+3)}P_{l+2}(a)+\left[\frac{(l+1)^2}{(2l+1)(2l+3)}+\frac{l^2}{(2l+1)(2l-1)}\right]P_l(a)\nonumber\\
& &+\frac{l(l-1)}{(2l+1)(2l-1)}P_{l-2}(a)\label{id_a2P},
\eea
together with the corresponding expression for $b^2 Q_l(b)$.

As in the electromagnetic case, the $l=0,1$ pieces of $S_2$ that do not multiply $S$ 
are chosen to cancel the $l=0,1$ pieces of the series, so that $S_2$ can be rewritten as
\bea
S_2&=&\sum_{l=2}^\infty(a^2+b^2+a^2 b^2-1-4b x+2x^2)(2l+1) P_l (a) Q_l(b) P_l(x)\\
&\equiv&\sum_{l=2}^\infty(A+B x+2x^2)(2l+1) P_l (a) Q_l(b) P_l(x).\label{S2b}
\eea
We use Eq.~(\ref{id_a2P}) (and the corresponding expression for $b^2Q_l(b)$) to write the part of Eq.~(\ref{S2b}) involving $A$ in the form

\bea
\sum_{l=2}^\infty &A&(2l+1) P_l (a) Q_l(b) P_l(x)\nonumber\\
&=&\sum_{l=2}^\infty \bm \left(\ \  \left\{\frac{(l+1)(l+2)}{(2l+3)}P_{l+2}(a)
+\left[\frac{(l+1)^2}{(2l+3)}+\frac{l^2}{(2l-1)}\right]P_l(a)\right.
+\frac{l(l-1)}{(2l-1)}P_{l-2}(a)\right\}Q_l(b)\nonumber\\
& &\ +P_l(a)\left\{\frac{(l+1)(l+2)}{(2l+3)}Q_{l+2}(b)
+\left[\frac{(l+1)^2}{(2l+3)}+\frac{l^2}{(2l-1)}\right]Q_l(b)
+\frac{l(l-1)}{(2l-1)}Q_{l-2}(b)\right\}\nonumber\\
& &\ +\frac1{2l+1}\left\{\frac{(l+1)(l+2)}{(2l+3)}P_{l+2}(a)+\left[\frac{(l+1)^2}{(2l+3)}+\frac{l^2}{(2l-1)}\right]P_l(a)+\frac{l(l-1)}{(2l-1)}P_{l-2}(a)\right\}\nonumber\\
& &\qquad\quad\times\left\{\frac{(l+1)(l+2)}{(2l+1)(2l+3)}Q_{l+2}(b)+\left[\frac{(l+1)^2}{(2l+1)(2l+3)}+\frac{l^2}{(2l+1)(2l-1)}\right]Q_l(b)\right.\nonumber\\
& &+\left.\frac{l(l-1)}{(2l+1)(2l-1)}Q_{l-2}(b)\right\}\ \left.\ -(2l+1)P_l(a)Q_l(b)\phantom{\frac AB}\!\!\!\bm\right) P_l(x).\label{AS}
\eea
We next use Eq.~(\ref{idxPl}) to rewrite $\sum_{l=2}^\infty B x (2l+1) P_l (a) Q_l(b) P_l(x)$ as
\bea
\sum_{l=2}^\infty &B& x (2l+1) P_l (a) Q_l(b) P_l(x)\nonumber\\
=&-&4\sum_{l=2}^\infty\Big\{l\left[\frac l{(2l-1)}P_l(a)+\frac{l-1}{2l-1}P_{l-2}(a)\right]\left[\frac l{(2l-1)}Q_l(b)+\frac{l-1}{2l-1}Q_{l-2}(b)\right]\nonumber\\
&+&\!\!(l+1)\!\!\left.\left[\frac {(l+2)}{(2l+3)}P_{l+2}(a)+\frac{l+1}{2l+3}P_l(a)\right]\!
\!\!\left[\frac {(l+2)}{(2l+3)}Q_{l+2}(b)+\frac{l+1}{2l+3}Q_l(b)\right]\!\!\right\}P_l(x).\qquad 
\label{BxS}
\eea
Substituting in Eq.~(\ref{S2b}) the expressions from Eqs.~(\ref{AS}), (\ref{BxS}), 
and (\ref{x2S}), we find 
\bea
S_2&=&\sum_{l=2}^\infty\frac{(l+2)(l+1)l(l-1)}{(2l-1)^2(2l+1)(2l+3)^2}\left[(2l-1)P_{l+2}(a)-2(2l+1)P_l(a)+(2l+3)P_{l-2}(a)\right]\nonumber\\
& &\phantom{xxxxxxxxxxxx}
\times\left[(2l-1)Q_{l+2}(b)-2(sl+1)Q_l(b)+(2l+3)Q_{l-2}(b)\right]P_l(x).
\label{S2_end}
\eea
From the definitions~(\ref{Plm}) and the identities (\ref{idPl}), we obtain
\bea
(a^2-1)P_l^2(a)=\frac{(l+2)(l+1)l(l-1)}{(2l-1)(2l+1)(2l+3)}&\Big[&\!\!\!(2l-1)P_{l+2}(a)\nonumber\\
&&-2(2l+1)P_l(a)+(2l+3)P_{l-2}(a)\Big],\label{Pl2_reln}
\eea
together with the corresponding equation for $Q_l^2(b)$.
Using these relations, Eq.~(\ref{S2_end}) takes the form (\ref{S2}), as claimed.

Finally, from Eqs.~(\ref{grsumpl}) and (\ref{S2}), with \dis a=\frac{{\mathfrak r}_<}M,\  
b=\frac{{\mathfrak r}_>}M,\ x=\cos\theta$ as before, we obtain the closed form expression 
(\ref{gr_psi4_closed}) for $\psi_4$:
\bea
\psi_4&=&-\frac{{\mathfrak m}}{4 M} \frac{\Delta\Delta_0^{1/2}}{r^4
r_0}\bar\eth^2\left[\frac M {\Delta\Delta_0}
\frac{M^2(\mathfrak r^2+\mathfrak r_0^2)+\mathfrak r^2\mathfrak
r_0^2-M^4-4\mathfrak r\mathfrak r_0
M^2\cos\theta+2M^4\cos^2\theta}{R}\right]\nonumber\\
&=& -{\mathfrak m}\,\frac34\frac{\Delta_0^{3/2}}{r_0}
                \frac{\Delta^2\sin^2\theta } {r^4 R^5}.
\eea

\end{document}